\documentclass{aa}  
\usepackage{xcolor}
\usepackage{subcaption}

\usepackage{graphicx}
\usepackage{txfonts}
\usepackage{amsmath}
\usepackage{longtable}

\begin{document}

   \title{Multi-spacecraft observations of the decay phase of solar energetic particle events}
\author{R.A. Hyndman\inst{\ref{inst1}}
\and S. Dalla\inst{\ref{inst1}}
\and T. Laitinen\inst{\ref{inst1}}
\and A. Hutchinson \inst{\ref{inst2}, \ref{inst3}} 
\and C.M.S. Cohen\inst{\ref{inst4}}
\and R.F. Wimmer-Schweingruber\inst{\ref{inst5}}}

\institute{Jeremiah Horrocks Institute, University of Central Lancashire, Preston, United Kingdom\\ \email{rahyndman@uclan.ac.uk}\label{inst1} 
\and
Heliophysics Science Division, NASA Goddard Space Flight Center, Greenbelt, MD 20771, USA\label{inst2}
\and
Goddard Planetary Heliophysics Institute, University of Maryland, Baltimore County, Baltimore, MD 21250, USA\label{inst3}
\and California Institute of Technology, Pasadena, CA 91125, USA;\label{inst4}
\and
Institute of Experimental and Applied Physics, Christian-Albrechts-University Kiel, Leibnizstraße 11, 24118 Kiel, Germany\label{inst5}
}
   \date{Received Month Day, Year; accepted Month Day, Year}
 
  \abstract
   { The parameters of solar energetic particle (SEP) event profiles such as the onset time and peak time have been researched extensively to obtain information on the acceleration and transport of SEPs. The corotation of particle-filled magnetic flux tubes with the Sun is generally thought to play a minor role in determining intensity profiles. However recent simulations have suggested that corotation affects the SEP decay phases and depends on the location of the observer with respect to the active region associated with the event.}
   {We aim to determine whether signatures of corotation are present in observations of the decay phases of SEP events, and we study the dependence of the parameters of the decay phase on the properties of the flares and coronal mass ejections associated with the events.}
   {We analysed multi-spacecraft observations of SEP intensity profiles from 11 events between 2020 and 2022 using data from Solar Orbiter, PSP, STEREO-A, and SOHO.
   We determined the decay-time constant, $\tau$, in three energy channels; electrons $\sim$1 MeV, protons $\sim$25 MeV, and protons $\sim$60 MeV. We studied the dependence of $\tau$ on the longitudinal separation, $\Delta \phi$, between the source of the active region and the spacecraft magnetic footpoint on the Sun.}
   {Individual events show a tendency for the decay-time constant to decrease with increasing $\Delta \phi$. This agrees with test particle simulations. The magnitude of the event as measured through the intensity of the associated flare and SEP peak flux affects the measured $\tau$ values and likely is the cause of the observed large inter-event variability together with the varying solar wind and the conditions in the interplanetary magnetic field.}
   {We conclude that corotation affects decay phase of an SEP event and should be included in future simulations and interpretations of these events.}

   \keywords{SEPs, 
                Sun: Corotation, 
                SEPs: Decay phases
               }

   \maketitle
%

\section{Introduction}\label{introduction}

Solar energetic particles (SEPs) are accelerated by shocks driven by coronal mass ejections (CMEs) and during flares in gradual and impulsive events respectively. They can then be detected as sporadic increases in the particle intensities up to and past 1 au from the Sun \citep{klein_acceleration_2017} by spacecraft in interplanetary space. Typical SEP time-intensity profiles have a rise phase, a peak intensity, and a decay phase, and gradual events have more complex profiles in some cases, in association with the passage of a shock at the spacecraft. The decay phases, from peak to background levels, can last between a few hours to several days \citep{van_allen_impulsive_1965}.

Historically, the analysis of SEP profiles has been based on one spacecraft observing many events from different source locations on the Sun. This allowed the investigation into the effect of the source location on the time profiles \citep{cane_role_1988}, but this type of study cannot separate differences caused by the observer location from those caused by the fact that events are produced by different solar eruptions, and that the solar wind and interplanetary magnetic field (IMF) conditions vary from one event to the next.

We currently are in a golden era of SEP research in which multiple spacecraft are capable of taking simultaneous SEP measurements in different locations in interplanetary space.
This allows us to compare profiles seen in different locations from the same source active region (AR), which reduced the effects from parameters that change from one event to the next. Multiple spacecraft around the Sun also allow us to observe particles more frequently because more spacecraft mean more opportunities to observe an event \citep{rodriguez-garcia_solar_2024}. 

The decay phases of SEP events were originally thought to be indicative of the turbulence-induced scattering experienced by SEPs in interplanetary space before they reach the observer. It was thought that longer decay phases resulted from stronger scattering conditions and shorter decay phases resulted from weaker scattering conditions. A value for the scattering mean free path was derived by fitting the intensity profile \citep{kallenrode_propagation_1992} as modelled via 1D focussed transport.
The gradual and impulsive scheme for events was described in the 1990s and linked the acceleration of SEPs to CME-driven shocks for gradual events. The duration of the decay phase for gradual events has since been thought to be related to time-extended shock acceleration, especially at low energies, such that a longer acceleration leads to a longer duration of the decay phase \citep{reames_spatial_1996}. 

In addition to turbulence and acceleration duration, solar rotation can also affect the temporal profile of an SEP event. The outward flow of the solar wind from the Sun generates magnetic flux tubes that are wound into the Parker spiral by the Sun's rotation. As the Sun spins, the magnetic flux tubes also rotate with the Sun from east to west. This effect is referred to as corotation \citep{mccracken_decay_1971}. From an intuitive point of view, some effects of corotation on SEP intensity profiles are expected. Neglecting significant cross-field diffusion, corotation causes particle-filled magnetic flux tubes to be `pulled' along westward over time, relative to an observer, after the SEPs are injected into space. This means that when an observer views an event from an eastern source region the filled flux tubes rotate towards them, and when an observer views a western event the filled flux tubes rotate away from them. This would affect the decay phase, with western events in particular being cut short. If corotation has a significant effect on the decay phases of SEP events, this east-west difference should be visible in comparisons of the decay-phase duration against the observer location.

Some studies have included corotation such as \citet{giacalone_longitudinal_2012} and \cite{laitinen_forecasting_2018}. \citet{giacalone_longitudinal_2012} simulated impulsive events and included corotation through the movement of field lines over time. They stated that the rotation of the field line, along with other transport effects, allows these compact events to be seen at wide longitudes. \citet{laitinen_forecasting_2018} used a simple 1D diffusion model to simulate the SEP propagation from a flare-like injection, and compared simulations with and without corotation. They concluded that corotation affects the event profiles, citing decay-phase and intensity differences at different longitudes with respect to the source ARs. \citet{daibog_characteristics_2006} used single-spacecraft observations to investigate the effects of the observer longitude on the proton intensity profiles. They concluded that the trend they observed was due to the rotational effect discussed in \cite{mccracken_decay_1971}.

However, in the study and modelling of gradual SEP events the role of corotation is usually neglected. This is based upon the results of 1D focussed transport models that included corotation in an approximate way \citep{lario_energetic_1998, kallenrode_propagation_1997} and concluded that it has negligible effects. \citet{reames_spatial_1997} also concluded that corotation does not play an important role by analysing a few SEP events with spatially and temporally invariant spectra.

Recent 3D test-particle modelling of SEPs injected by a wide shock-like source has suggested that corotation has a significant effect on the decay phases of SEP events. \cite{hutchinson_impact_2023} ran simulations of SEP propagation that modelled particle transport with and without corotation. They found that including corotation had a notable effect on the decay phases, with the decay-time constant $\tau$ displaying a dependence on the longitudinal separation between source AR and observer footpoint (their Figure 3).

\cite{lario_statistical_2010} used single-spacecraft observations to investigate the decay phases of near relativistic electron events. Figure 10 in their paper analysed the dependence of $\tau$ on the source AR longitude. They found no dependence of $\tau$ on the source AR longitude in their dataset.

This work aims to investigate SEP decay phases during 11 multi-spacecraft events. We use data from four spacecraft: Solar Orbiter (SolO), the Parker Solar Probe (PSP), the Solar and Heliospheric Observatory (SOHO), and the Solar TErrestrial RElations Observatory - Ahead (STEREO-A). We analyse data for protons and electrons and fit the intensity profiles to obtain the value of the decay-time constant $\tau$. We analyse the dependence of $\tau$ on the relative location between observers and the source AR of the event, as well as the parameters of the solar events accelerating the particles. We investigate possible signatures of corotation and compare any results to those of the simulations of \cite{hutchinson_impact_2023}.

In Section 2 we discuss the method we used to find decay-time constant values for the measured intensity-time profiles. In Section 3 we present our results and compare $\tau$ values with the locations of the observing spacecraft relative to the AR and the parameters of each event. In Section 4 we present our conclusions.

\section{Data and method}\label{methods} 
\subsection{Data sources}

We used data from Solar Orbiter \citep{muller_solar_2020}, PSP \citep{fox_solar_2016}, SOHO \citep{domingo_soho_1995} and STEREO-A \citep{kaiser_stereo_2008} to create SEP multi-spacecraft intensity-time plots. The instruments used for each are the Solar Orbiter High Energy Telescope (HET) \citep{rodriguez-pacheco_energetic_2020}, PSP Integrated Science Investigation of the Sun Energetic Particle Instrument-High High-Energy Telescope  (IS$\odot$IS/EPI-Hi/HETA) \citep{mccomas_integrated_2016}, the SOHO Energetic and Relativistic Nuclei and Electron experiment (ERNE) \citep{torsti_energetic_1995, valtonen_energetic_1997} and the Electron Proton Helium INstrument (EPHIN) \citep{muller-mellin_costep_1995}, and the STEREO-A High Energy Telescope (HET) \citep{von_rosenvinge_high_2008}.

We chose to use multiple energy channels to determine any differences in our results based on particle species or energy. The energy channels of the different instruments do not overlap exactly. For our multi-spacecraft analysis we identified channels with similar ranges for protons around 25 MeV and 60 MeV (see Table \ref{tab:Channels}). For electrons we used channels with ranges around 1 MeV. In Table \ref{tab:Channels} we list the relevant spacecraft instruments and energy channels for different particle species. The particle intensities are given in standard units for all spacecraft and channels except for the PSP/IS$\odot$IS electron $\sim$ 1 MeV channel, where count-rate data were used. The count-rate data for the PSP channels that cover the range between 0.6 and 1.2 MeV were summed to obtain a 0.6 - 1.2 MeV channel that is more comparable to the channel widths of the other instruments. This channel alone was formed from several summed channels. All data were downloaded using the SERPENTINE analysis tools \citep{palmroos_solar_2022}. SOHO, STEREO-A, and Solar Orbiter data have a 30-minute cadence, and PSP data are available at a 60-minute cadence.

\subsection{Event selection}
We selected events between 2020 and 2022 based on the availability of multi-spacecraft data. An event was selected for study when:
\begin{itemize}
  \item at least two spacecraft had observed the event in at least one of the energy channels we used
  \item the observing spacecraft were at a radial distance was farther than 0.6 au from the Sun. This was to ensure that only a relatively small range of distances were used to ensure similar transport conditions, and to avoid events close to the Sun, where the spacecraft are moving fast across longitudes, which may influence the profiles
  \item the observations had reliable count-rate and intensity-time data. This required the observations to have no significant gaps in the data, such that fits to the decay phase were within the goodness-of-fit requirements discussed in Section \ref{DTC}
  \item any events occurring before or after the chosen event could be separated from the decay phase of the event being studied.
\end{itemize}

The 11 selected events are listed in Table \ref{tab:Events}. Further details for the events, including the parameters we calculated in our study, can be found in the appendix in Table \ref{tab:long}.


\begin{table*}[h]
\caption{Energy channels for the multi-spacecraft analysis.}
\begin{center}
\small

\begin{tabular}{ |*{6}{c|} }
    \hline 
   Instrument &  SolO HET & PSP IS$\odot$IS & SOHO ERNE & SOHO EPHIN & STEREO-A HET \\ 
    \hline 
 Electrons  $\sim$ 1 MeV & 1.05-2.40 & 0.6-0.7& N/A & 0.67-10.4 & 0.7-1.4 \\
&   &  0.7-0.8  &   &   &   \\
&   &  0.8-1.0   &   &   &   \\
&   &  1.0-1.2   &   &   &   \\
    \hline 
 Protons $\sim$ 25 MeV & 25.09-27.20 & 26.9-32.0 & 25-32 & 25-53 & 26.3-29.7 \\ 
    \hline 
 Protons $\sim$ 60 MeV & 63.10-68.97   & N/A  & 64-80  & N/A  & 60-100 \\
    \hline 
\end{tabular}
\tablefoot{For electrons, count rates in the 4 PSP channels shown were summed.}
\end{center}

\label{tab:Channels}

\end{table*}

\begin{table*}[]
\caption{Associated flare and CME properties for the SEP events.} 
\begin{center}
\small
\begin{tabular}{ |c||c|c|c|c|c| } 
    \hline 
     Event Date & \shortstack{Flare Time\\(UTC)} & \shortstack{Flare Location \\ (Stonyhurst, [lon., lat.], \textdegree)} & Flare Class &  CME Speed $\rm{( km s^{-1})}$ & \shortstack{\\Peak Flux, \\ $I_{p,25}^{max}$ $\rm{( cm^{-2}s^{-1}sr^{-1}MeV^{-1})}$ } \\ 
    \hline 
    29-11-20 & 12:34:00 & [-82.0, -23.0] & M4.4 & 2077 & 4.64e+00 \\
     \hline
    07-12-20 & 15:46:00 & [8.2, -25.0] & C7.4 & 1407 & 7.11e-02 \\ 
     \hline
    28-05-21 & 22:19:00 & [54.9, 19.0] & C9.4 & 971 & 2.35e-01 \\ 
     \hline
    09-06-21 & 11:50:00 & [89.0, 27.0] & C1.7 & 441 & 9.28e-02 \\ 
     \hline
    09-10-21 & 06:19:00 & [-8.3, 18.0] & M1.6 & 712 & 3.32e-01 \\ 
     \hline
    28-10-21 & 15:17:00 & [1.2, -28.0] & X1.0 & 1519 & 5.13e+01 \\ 
     \hline
    20-01-22 & 05:41:00 & [75.8, 8.0] & M5.5 & 1431 & 2.45e-01 \\ 
     \hline

    15-02-22 & 21:50 UT* & [-134.0, 33.0]*  & - & 1905 & 7.13e+00 \\
     \hline
    14-03-22 & 17:13:36 & [109.0, -24.0] & B8.5 & 740 & 9.04e-02 \\ 
     \hline
    28-03-22 & 10:58:00 & [4.3, 14.0] & M4.0 & 905 & 6.65e-01 \\ 
     \hline
    11-05-22 & 18:08:00 & [89.3, -17.0] & M2.7 & 1100 & 5.19e-02 \\ 
    \hline 
\end{tabular}
\tablefoot{Event date and flare onset time and location are given. These times and locations are taken from the SERPENTINE Events Catalog \citep{dresing_serpentine_2024}. Entries marked with * are taken from \citet{khoo_multispacecraft_2024} rather than the SERPENTINE catalogue due to data availability. Event magnitude proxies such as flare class, CME speed and maximum peak flux, $I_{p,25}^{max}$, values are given. Flare classes and $I_{p,25}^{max}$ values are also taken from SERPENTINE. $I_{p,25}^{max}$ values are the maximum proton peak flux for the $\sim$ 25 MeV channel, across the 4 observing spacecraft we use. CME speeds are plane-of-sky speeds from LASCO. Where more than one CME is associated, the largest plane-of-sky LASCO value is used.}
\end{center}

\label{tab:Events}
\end{table*}

\subsection{Decay-time constant}\label{DTC}

In order to quantify the decay phase of the SEP events, we defined the decay-time, $\tau$, as

\begin{equation}\label{eqn:tauEqn} 
    \tau = -  \left( \frac{\mathrm{d}(\ln I)}{\mathrm{d}t} \right) ^{-1}
\end{equation}
where $I$ is the particle intensity, and $t$ is the time.

To derive $\tau$, we started by removing the background intensities. We averaged the background intensities over 20 hours, from one day to 4 hours prior to the event start time.
When this time frame included previous events, we instead selected a 20-hour time frame when the intensities were at background level before all the events. We took the mean of the background intensities during these times and subtracted the mean from the data from each instrument.

We then used scipy.stats.linregress\footnote{https://docs.scipy.org/doc/scipy/reference/generated/scipy.stats.linregress.html} to fit a straight line to the decay phase on logarithmic intensity-time plots.
The period to fit the decay phase started from the time when the intensity value fell to 90\% of the peak intensity, $I_{p}$, and it ended when particle intensities again reached background levels (or until a following event was detected). We fitted from 90\% of the $I_{p}$ value instead of from $I_{p}$ itself, to prevent a too early start of the decay phase we measured. In some profiles, the intensities plateaued at the $I_{p}$ for a time before the decay set in which required the intensity to drop by 10\% and allowed us to obtain a more accurate decay-phase duration. We combined four channels to derive a synthetic channel at energy 0.6-1.2 MeV for PSP electrons and we therefore verified the values of the decay-time constant for the four individual channels. We found that the decay-time constant was very similar for each of the four channels.

To evaluate the goodness-of-fit for each decay slope, we used the p value returned by scipy.stats.linregress. This is the p value for a hypothesis test using the Wald test where the slope of the regression line is lower than zero. 
A fit was accepted when the p value of the fit was $\leq$ 0.05 and the fit lasted 9 hours at least. The minimum time-limit was set to allow for a significant number of points for the analysis because most data had at a 30 minute cadence (Solar Orbiter, SOHO, and STEREO data). Fits with more than 18 points were judged to be accurate using the goodness-of-fit p value and also by eye. Eighteen points were also required for the PSP data, although this corresponds to 18 hours as these were 60-minute-averaged data. Previous studies of the decay phase also placed time limits on decay phases to be included in analysis. For example \citet{daibog_characteristics_2006}  took 12 hours as their lower limit. After fitting a straight line to the decays, the gradients of the straight lines were converted to the decay-time constant values, $\tau$,  using Equation \ref{eqn:tauEqn}.

The error bars were calculated from values given by the fitting routine. The maximum and minimum values for the decay-phase fit were converted into hours using Equation \ref{eqn:tauEqn} and were added to each point. The maximum slope value corresponds to the lowest $\tau$ value, and the minimum slope value corresponds to the highest $\tau$ value. 

For some of the SEP intensity profiles, the decay seems to have two phases. \cite{lario_statistical_2010} noted that some SEP events have two decay phases where the earlier stage follows a power law and the later stage follows an exponential decay. \cite{lario_statistical_2010} found that observations of one-phase decays were mostly found outside the nominal well-connected longitudes. 
To verify whether two-phase decays affects the $\tau$ values significantly, we also ran our fitting routine starting the fit when the intensities reached 50\% of $I_{p}$ to background. This allowed us to obtain $\tau$ values that only focussed on late decay. This is discussed further in Section \ref{LarioDecay}.

\subsection{Solar event properties}

Source ARs of the solar events associated with the SEP event were obtained from the SERPENTINE event catalogue \citep{dresing_serpentine_2024}. The longitude values of the ARs were taken from the catalogue, as were the associated CME speeds, flare classes, and maximum peak particle flux for the $\sim$ 25 MeV proton channel, $I_{p,25}^{max}$, for the events (the CME speed and flare class values in the catalogue originally come from LASCO and GOES, respectively). 

To characterise the observer location with respect to the event source AR, we defined $\Delta \phi$, the difference in longitude between the observer's magnetic footpoint on the Sun and the AR associated with the event,

\begin{equation}\label{eqn:DPEqn}
    \Delta \phi = \phi_{AR} - \phi_{ftpt}
\end{equation}
where $\phi_{AR}$ is the longitude of the source AR, and  $\phi_{ftpt}$ is the longitude of the observer's magnetic footpoint on the Sun. 
We determined the $\phi_{ftpt}$ for each spacecraft using the open-source tool Solar Magnetic Connection HAUS (Solar-MACH) \citep{gieseler_solar-mach_2023}. $\phi_{ftpt}$ was calculated using the nominal Parker Spiral for each spacecraft based on the solar wind speed, $V_{sw}$. $V_{sw}$ was taken from Coordinated Data Analysis Web (CDAWeb)\footnote{https://cdaweb.gsfc.nasa.gov}. The $V_{sw}$ value observed closest to the event start time, and within two hours of the start time was used. The event start times were taken from the SERPENTINE \cite{events_catalog_serpentine_serpentine_2024}.
When $V_{sw}$ data were not available within two hours of the event start time, we used 450 $\rm{kms^{-1}}$ as a default value. When the AR lay west of the footpoint of the spacecraft (western events), $\Delta \phi >$ 0, and when the AR lay east of the footpoint of the spacecraft (eastern events), $\Delta \phi <$ 0. 

\section{Results}\label{results} 

\begin{figure}
    \centering
    \includegraphics[width=1\linewidth]{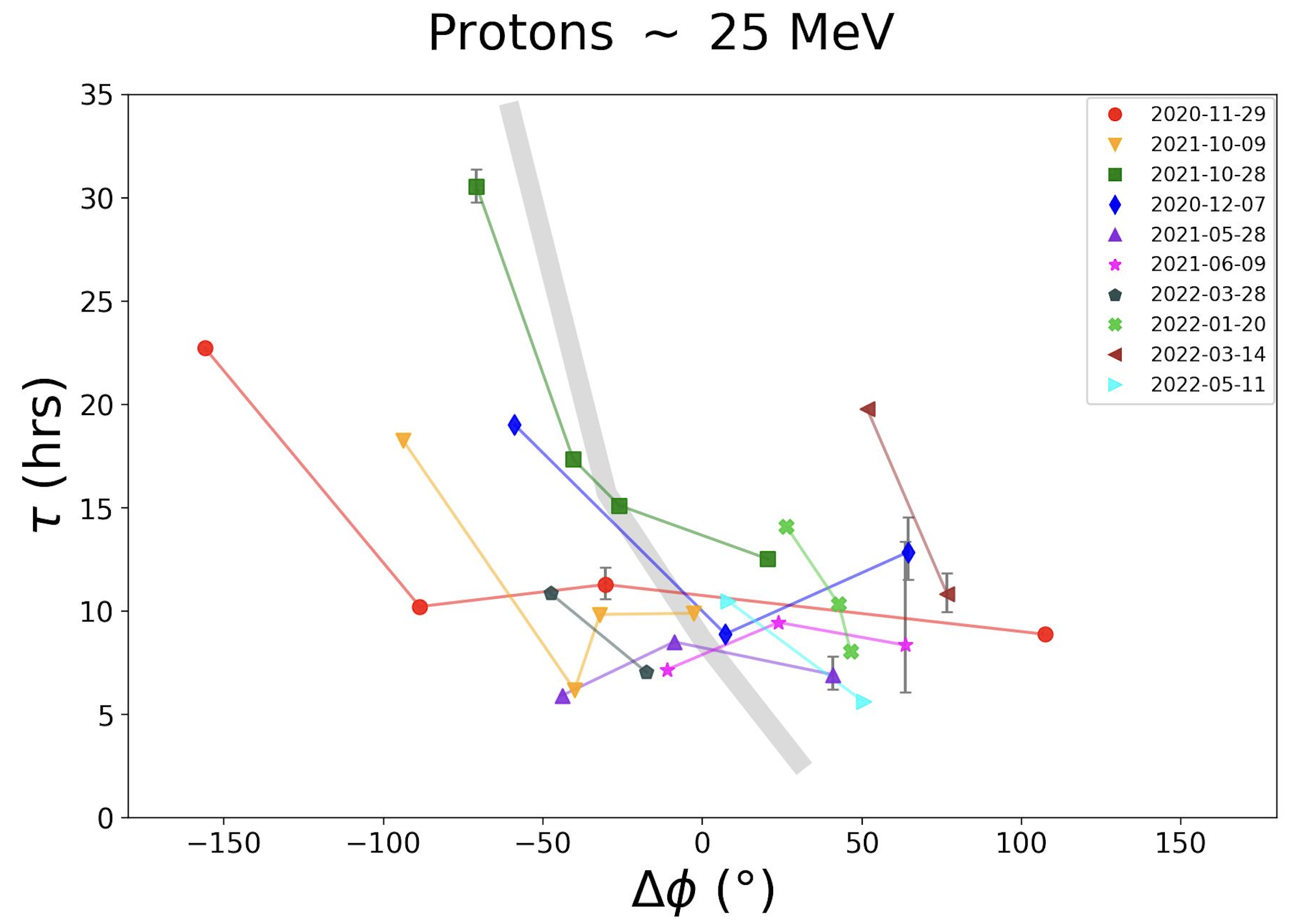}
    \caption{Decay-time constant $\tau$ vs. longitudinal separation $\Delta \phi$ (as given by Equation \ref{eqn:DPEqn}) for protons at $\sim$ 25 MeV. The coloured lines connect s/c data points for a single event. The grey shading shows results from a 25 MeV simulation run following the method of \citet{hutchinson_modelling_2023}. The error bars are omitted when they are smaller than the data points.}
    \label{fig:P25TauPhi}
\end{figure}

\subsection{Decay-time constant against longitudinal separation}

In Figure \ref{fig:P25TauPhi} we plot the decay-time constant, $\tau$, against $\Delta \phi$ for $\sim$ 25 MeV protons for ten events. The 2022 February 15 event was not included for the reasons discussed in Section \ref{WideEvent}. The dependence was obtained by using similar simulations to those by \citet{hutchinson_impact_2023, hutchinson_modelling_2023} who ran 3D test particle simulations of SEPs injected from a shock-like source and propagated a mono-energetic (5 MeV) proton population for 72 hours. We ran 3D test-particle simulations for 25 MeV protons and kept all other parameters the same as in \citet{hutchinson_modelling_2023}. From this, we obtained the grey shaded area in Figure \ref{fig:P25TauPhi}.

Figure \ref{fig:P25TauPhi} shows that for a given $\Delta \phi$ value, a broad range of $\tau$ values are observed. In general, a larger spread and higher $\tau$ values are seen for eastern events ($\Delta \phi$ < 0 ) compared to western events ($\Delta \phi$ > 0). When we consider each event individually, there is a tendency for $\tau$ to decrease from east to west for most of the events. This agrees with the trend from the test-particle simulations.

\begin{figure}
    \centering
    \includegraphics[width=1\linewidth]{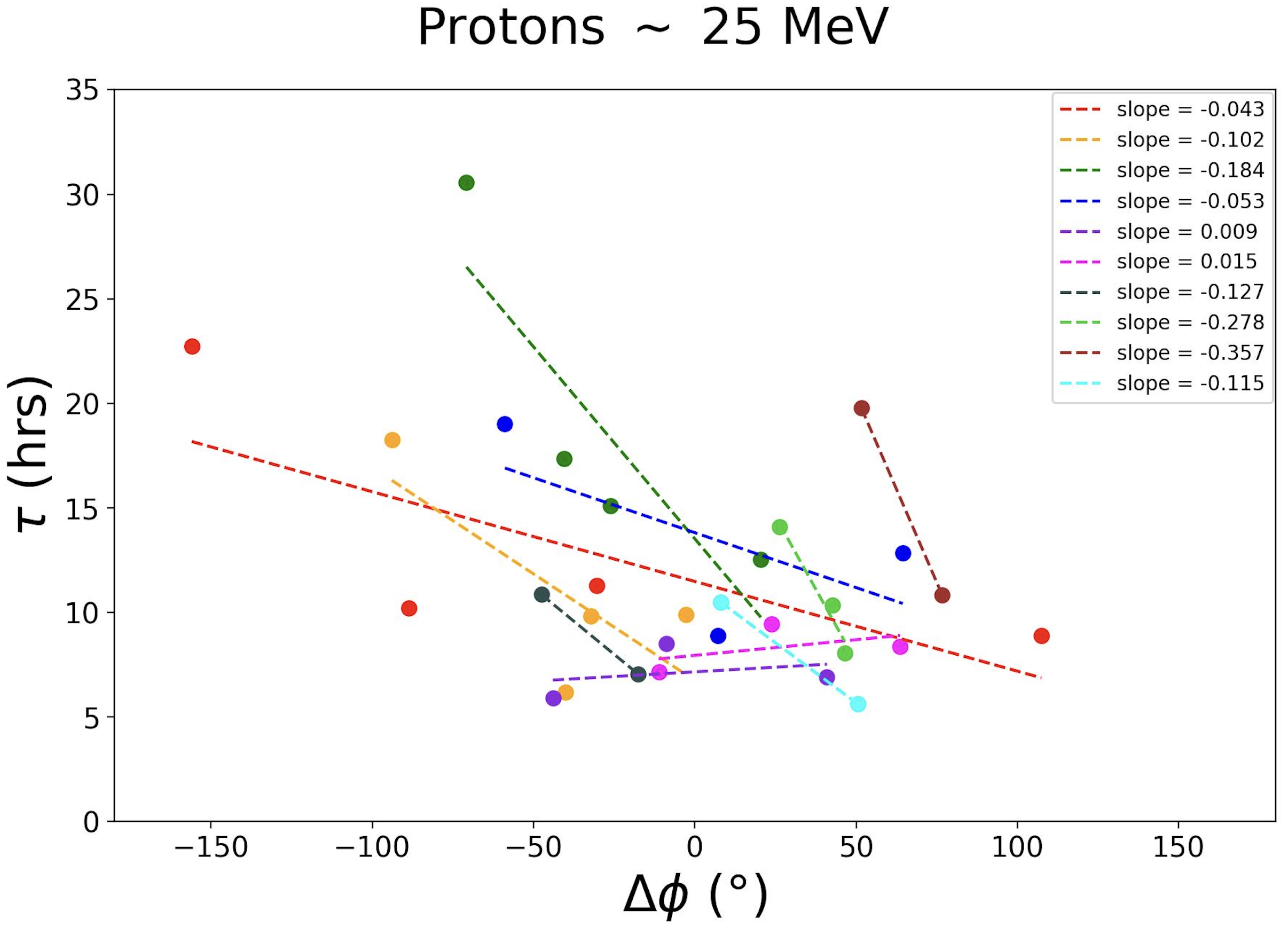}
    \caption{$\tau$ against $\Delta \phi$ with each event in a different colour. The best-fit dashed lines are plotted for each event and the gradient values ($hrs / ^{\circ}$) of these lines are shown in the key.}
    \label{fig:EvSlopes}
\end{figure}

\begin{figure}
    \centering
    \includegraphics[width=1\linewidth]{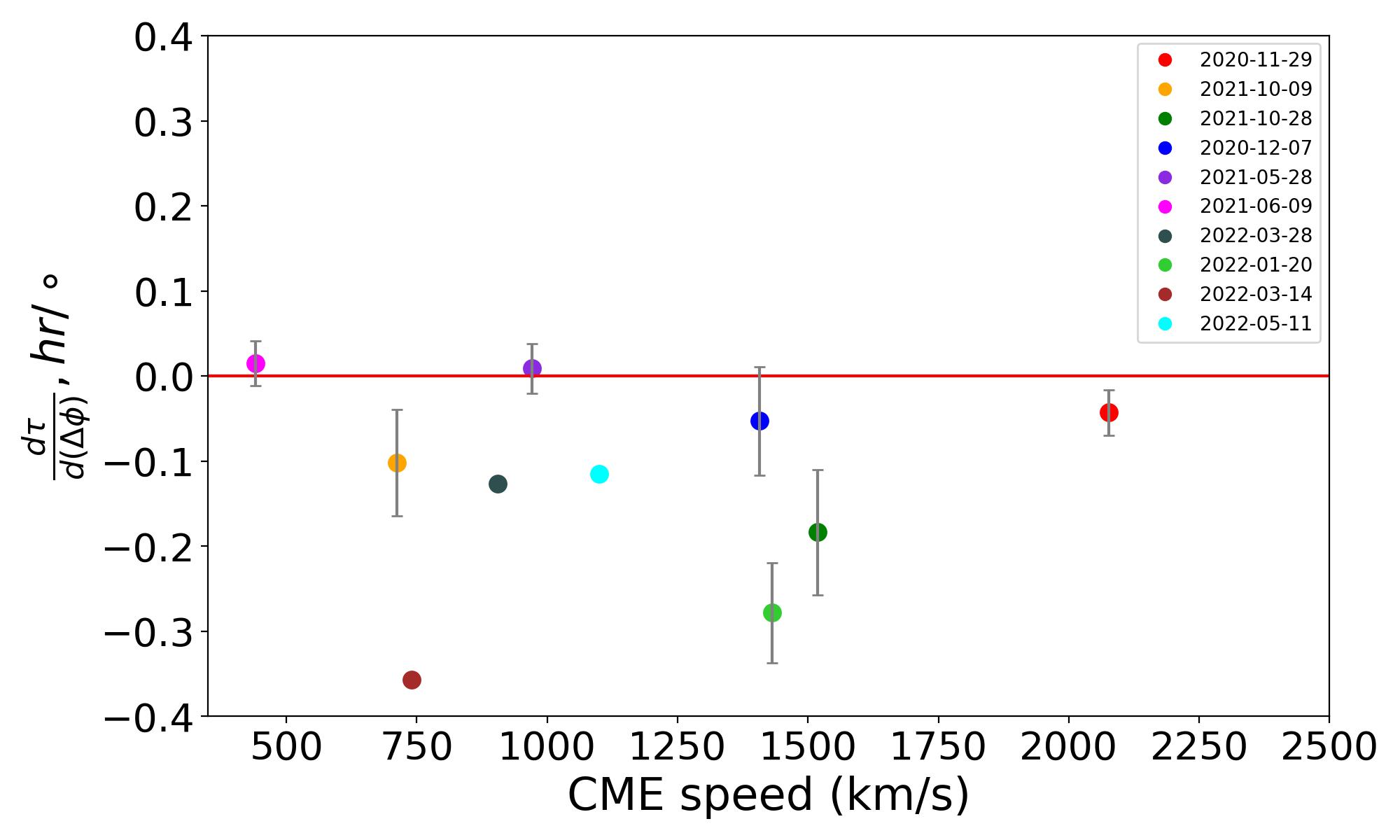}
    \caption{$\frac{\mathrm{d} \tau}{\mathrm{d}(\Delta \phi)}$ values from the linear fit of $\tau$ vs $\Delta \phi$ data points for each event against CME speed. The red line marks the gradient = 0 line.}
    \label{fig:SlopeCME}
\end{figure}

To capture the east-west trend, we show in Figure \ref{fig:EvSlopes} the best linear fit to the data points, which corresponds to an individual event for the proton $\sim$ 25 MeV channel. We call the gradient of these fits $\frac{\mathrm{d} \tau}{\mathrm{d}(\Delta \phi)}$, and the fits are shown as colour-coded dashed lines in Figure \ref{fig:EvSlopes}.

We plot $\frac{\mathrm{d} \tau}{\mathrm{d}(\Delta \phi)}$ of each event against CME speed in Figure \ref{fig:SlopeCME}. Almost all events have negative gradients, which supports the hypothesis of higher $\tau$ values in the east and lower $\tau$ values in the west. Two events have positive gradients, but both of these values are low. On the whole, a trend for higher eastern $\tau$ values and lower western $\tau$ values is seen. 

\begin{figure}
    \centering
    \includegraphics[width=0.95\linewidth]{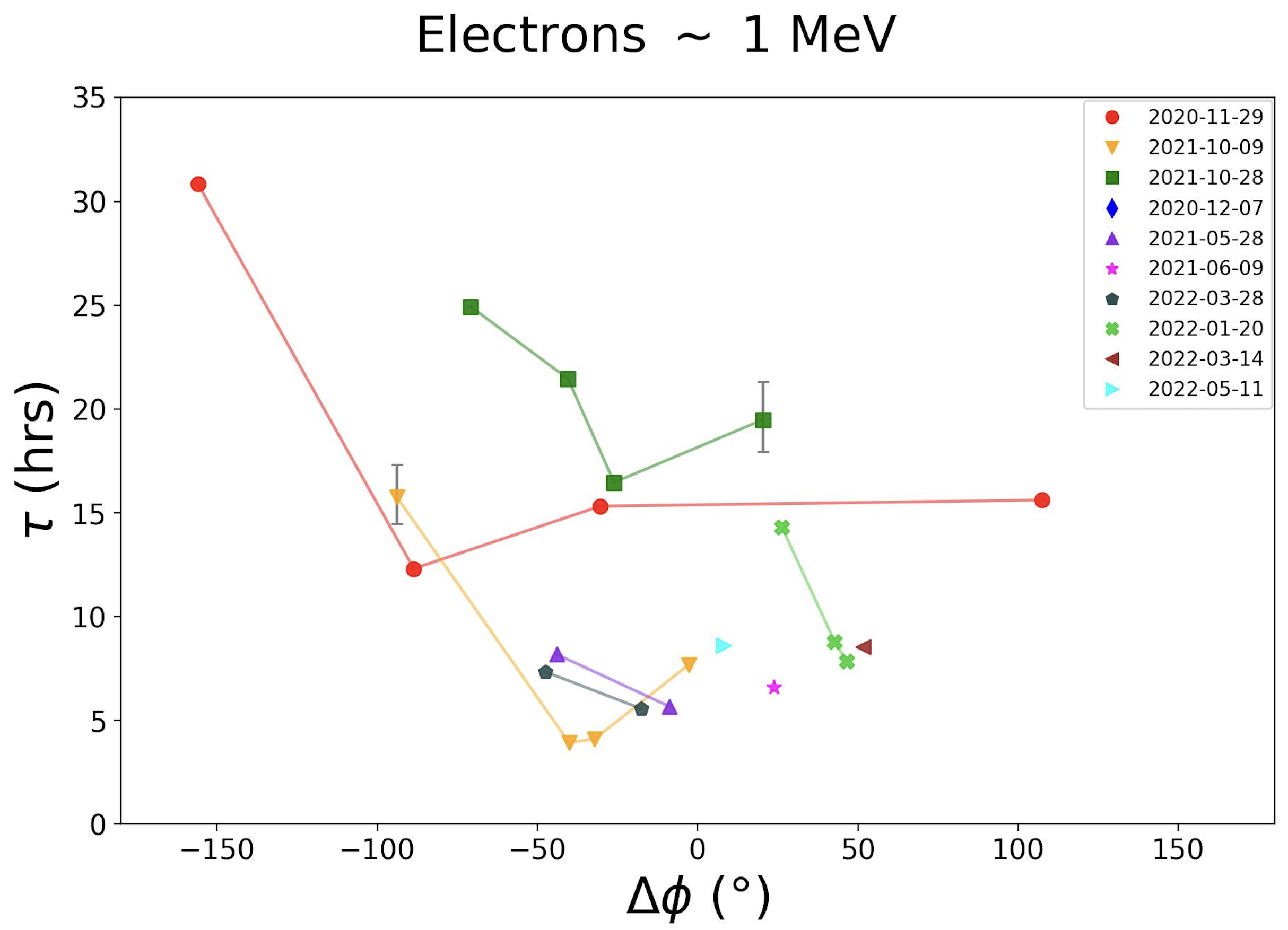}
    \caption{Decay-time constant $\tau$ vs. longitudinal separation $\Delta \phi$ (as given by Equation \ref{eqn:DPEqn}) for electrons at $\sim$ 1 MeV. The coloured lines connect s/c data points for a single event. The error bars are omitted when they are smaller than the data points.}
    \label{fig:E1TauPhi}
\end{figure}

\begin{figure}
    \centering
    \includegraphics[width=0.95\linewidth]{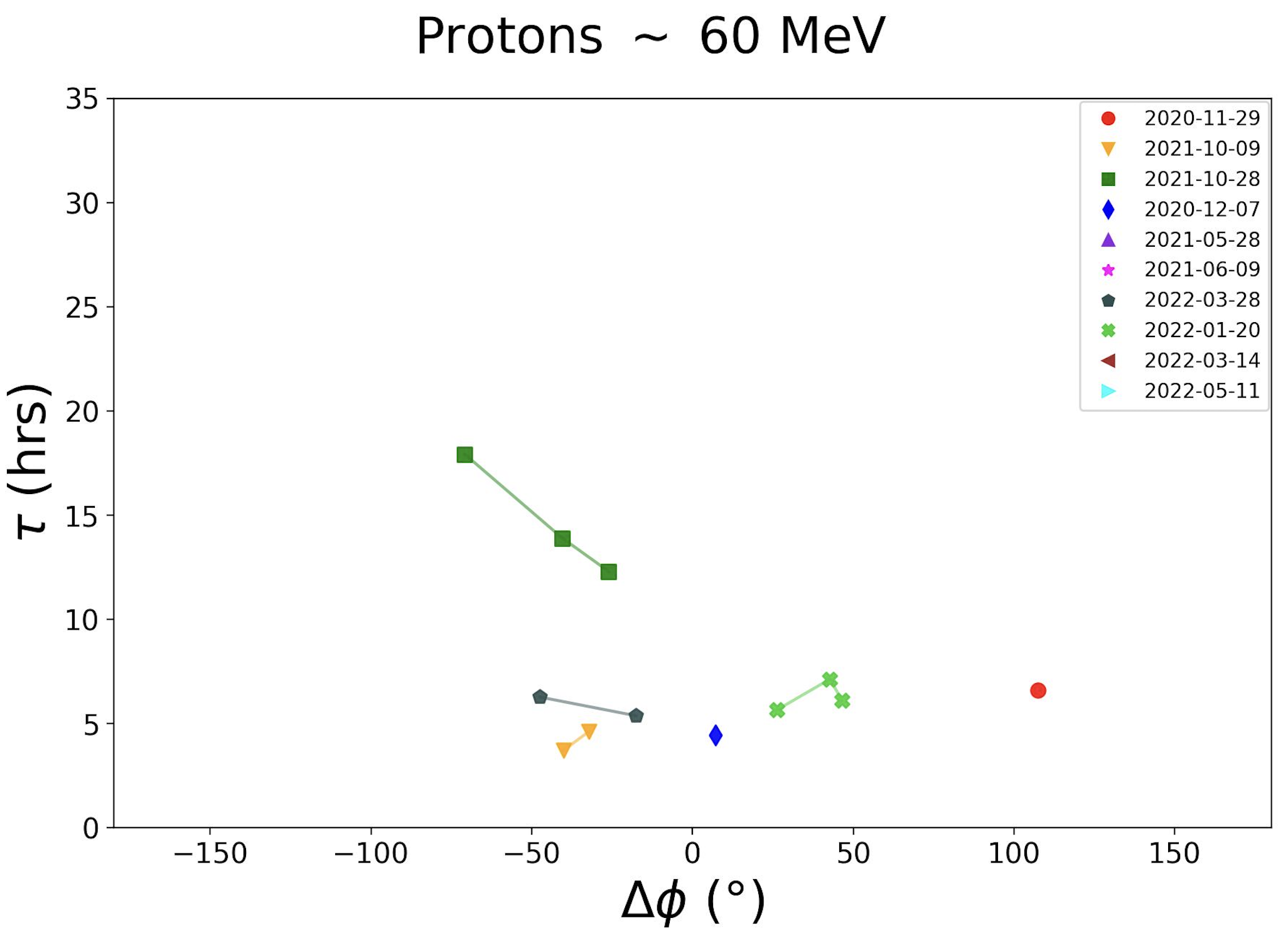}
        \caption{Decay-time constant $\tau$ vs. longitudinal separation $\Delta \phi$ (as given by Equation \ref{eqn:DPEqn}) for protons at $\sim$ 60 MeV. The coloured lines connect s/c data points for a single event. The error bars are omitted when they are smaller than the data points.}
    \label{fig:P60TauPhi}
\end{figure}

Our focus lay on the $\sim$ 25 MeV proton channel because it has the largest number of measurements. The data for other channels were also analysed, however. We plot the decay-time constant versus $\Delta \phi$ plots for $\sim$ 1 MeV electrons and $\sim$ 60 MeV protons in Figures \ref{fig:E1TauPhi} and \ref{fig:P60TauPhi} respectively. The $\sim$ 1 MeV electron channel and the $\sim$ 60 MeV proton channel display similar trends to those seen for $\sim$ 25 MeV protons, with fewer data points.

Figure \ref{fig:E1TauPhi} shows that the $\sim$ 1 MeV electrons have higher $\tau$ values than the $\sim$ 25 MeV protons for some events and lower values for others, which results in a larger spread in values for this particle species, but the general east-to-west decrease trend for each event is maintained. This might mean that corotation affects electrons to a different degree that it affects protons, although it  might also indicate other factors that may affect $\tau$ values for electrons differently to protons. These other factors are discussed in Section \ref{CaseStudySection}.

The values of $\tau$ for the $\sim$ 60 MeV proton channel tend to be slightly lower than those for the $\sim$ 25 MeV channel for the same observer, and event, indicating a faster decay at higher energy. In the $\sim$ 60 MeV proton channel, far fewer data points are available than in the other channels. One reason for this is that intensities are lower in this higher-energy channel, which means that decay phases are more difficult to fit with adequate statistics. Lower peak intensities also mean that background levels are reached earlier and the decay phase is cut off. No $\sim$ 60 MeV proton data are available from the PSP IS$\odot$IS/EPI-Hi instrument, and fewer measurements are therefore available at this energy. Another reason for fewer data points is that lower-magnitude solar events (as measured with proxies such as flare class, CME speed, and SEP peak flux) do not accelerate particles to this energy. Higher-magnitude events may accelerate particles to higher energies, but these high-energy particles may decelerate or escape to greater radial distances than the spacecraft location before they reach a spacecraft at large $| \Delta \phi|$. Alternatively, the flanks of the shock may not accelerate particles as efficiently as the nose of the shock at higher energies, resulting in fewer events. All these reasons may explain the lack of points at large $ | \Delta \phi | $ in Figure \ref{fig:P60TauPhi}.

\subsection{Comparing two events with similar geometries}\label{CaseStudySection}

\begin{figure*}
    \centering
    \includegraphics[width=.40\textwidth]{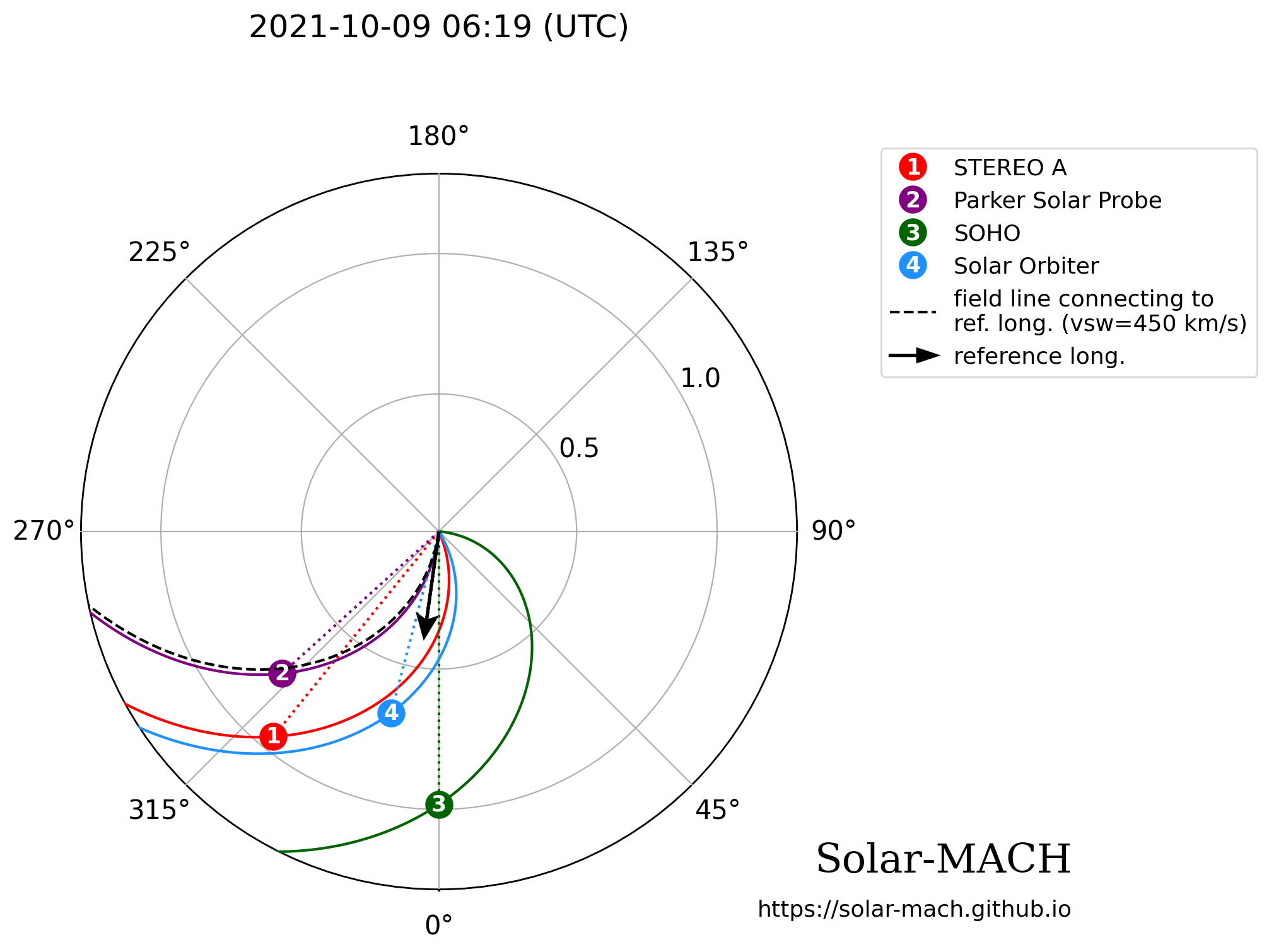}
    \includegraphics[width=.40\textwidth]{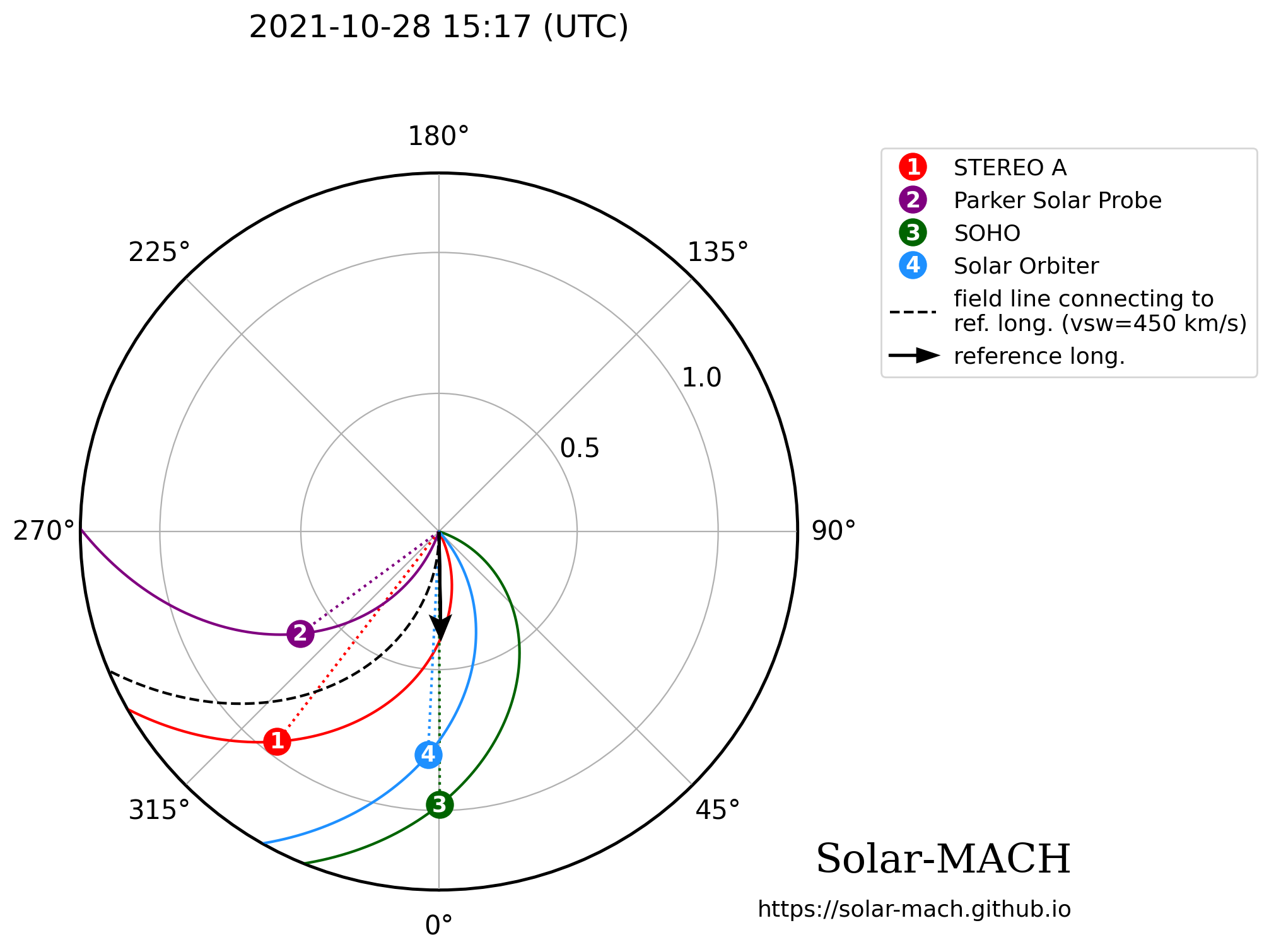}
    \includegraphics[width=.40\textwidth]{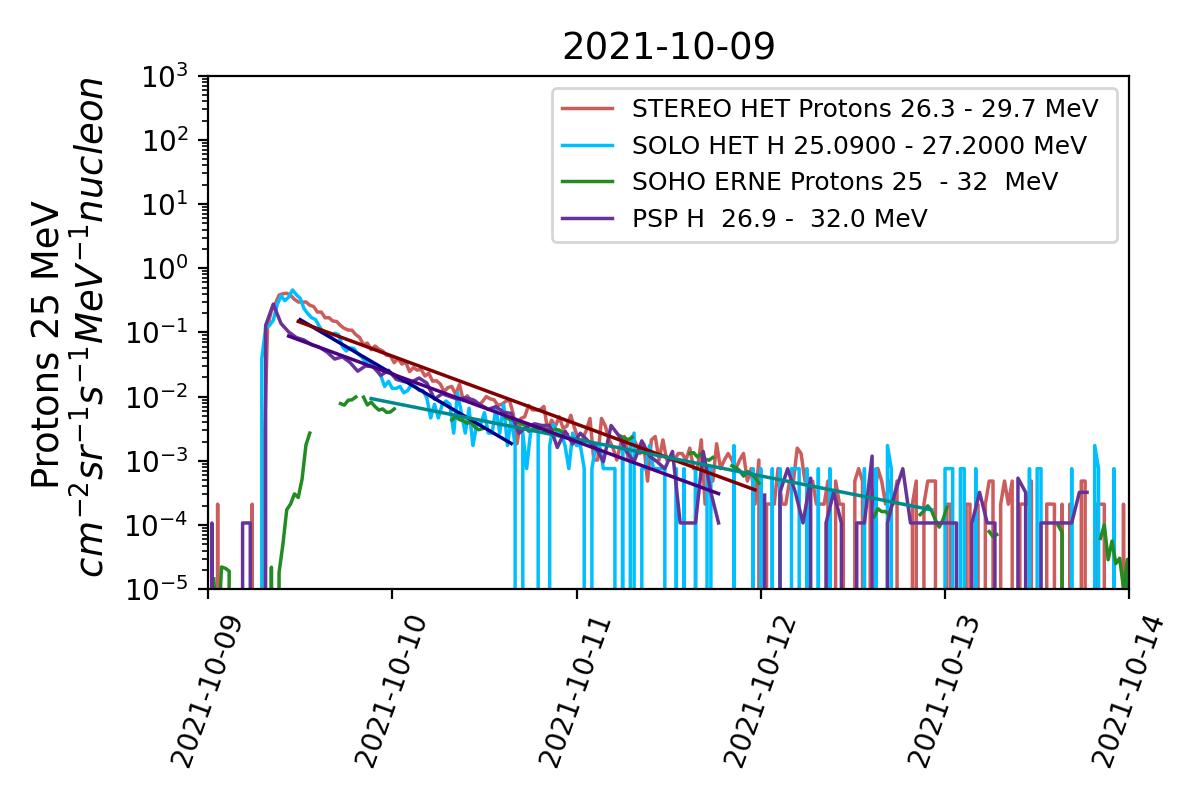}
    \includegraphics[width=.40\textwidth]{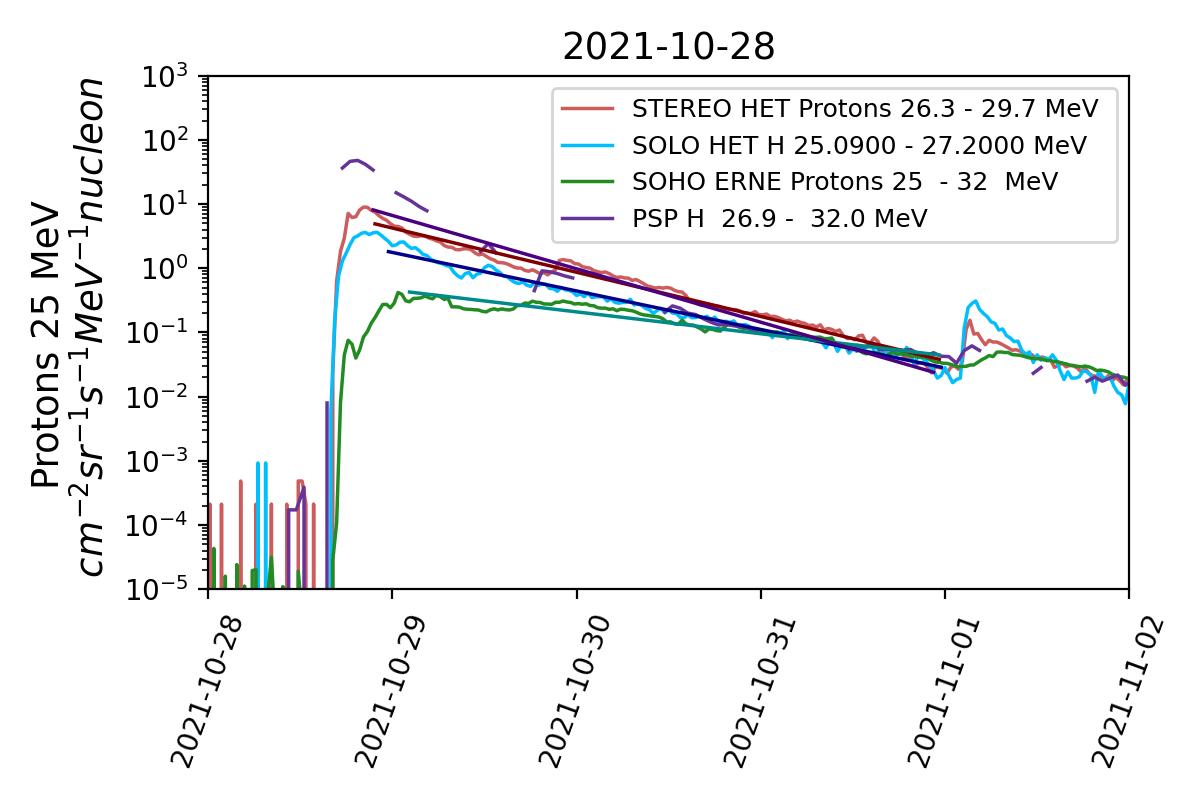}
    \caption{Spacecraft configurations (top panels) and SEP intensity profiles (bottom panels) for the two case study events 2021 October 9 (left) and 2021 October 28 (right). 
    The spacecraft configurations are from Solar-MACH. The colour-coded circles indicate the spacecraft locations and nominal Parker spirals calculated from measured solar wind speed at the spacecraft. 
    Multi-spacecraft SEP intensity profiles for the proton $\sim$ 25 MeV channel were made using SERPENTINE tools.
    The 2021 October 28 event decay is cut off by a second event that occurred on 2021 November 1.}
    \label{fig:SMCaseStudies}
\end{figure*}

A key characteristic that plays a role in determining the value of $\tau$ is the overall magnitude of the solar eruptive event that accelerated the particles. An example of this is presented in the following case study of two events with very similar geometries.

The events of 2021 October 9 and 2021 October 28 took place less than a month apart and the observing spacecraft moved very little in that time. Therefore, the locations of the spacecraft are very similar for the two events (see Figure \ref{fig:SMCaseStudies} top panels). Combined with the fact that the locations of the source ARs are close to each other in longitude (2021 October 9: E08N18 (Stonyhurst), 2021 October 28: W01S28), this means that the $\Delta \phi$ values are very similar.

If the geometrical locations were the only influence on $\tau$ value, we would expect the events to have similar values of $\tau$ for each spacecraft. However, as shown in Table \ref{tab:CaseStudyTable}, which lists the $\tau$ values for each spacecraft and channel for the two events, the $\tau$ values for the 2021 October 28 event are much higher than those for the 2021 October 9 event. We see this also in the comparison of the events in Figure \ref{fig:P25TauPhi} (yellow triangles and green squares).

The reason for these differing $\tau$ values may lie in the difference in magnitude of the events. 2021 October 9 is an M1.6 class flare event, while 2021 October 28 is an X1.0 class flare event. The events also have differing CME speeds, with 2021 October 9 having an associated CME plane-of-sky (POS) speed of 712 km/s while 2021 October 28 has a CME POS speed of 1519 km/s. The SEP peak flux for protons $\sim$ 25 MeV, $I_{p,25}^{max}$, for the events also differs, at around $3.32 \times 10^{-1}$ \textrm{cm}$^{-2}$\textrm{s}$^{-1}$\textrm{sr}$^{-1}$\textrm{MeV}$^{-1}$ for 2021 October 9, and the $5.13\times 10^{1} $ \textrm{cm}$^{-2}$\textrm{s}$^{-1}$\textrm{sr}$^{-1}$\textrm{MeV}$^{-1}$ for 2021 October 28. All three of these parameters can be used as proxies for the magnitude of the events. Overall, 2021 October 9 is much less energetic than the 2021 October 28 event. 

More energetic solar events are capable of accelerating particles up to higher energies than less energetic events, and there are many more particles at the high energies in more energetic events, and they are therefore visible above the background for a longer time. This could result in extended decays, in particular for lower-energy channels, which are caused by the deceleration of particles with higher energies that eventually fill these lower channels. More energetic events may also accelerate more particles over longer times and cause extended decay phases. The 2021 October 28 event was able to accelerate particles to much higher energies than the 2021 October 9 event, as demonstrated by the detection of an associated ground-level enhancement (GLE) event, showing that it accelerated protons to energies > 500 MeV.

In addition, more intense events tend to fill a broader region of the heliosphere with particles. Together with the corotation effect, this would result in longer decay phases for more intense events. We conclude that the higher $\tau$ values in the 2021 October 28 event are a result of the overall far higher event magnitude than for the 2021 October 9 event. Thus, the observed large variation in the $\tau$ values seen in Figure \ref{fig:P25TauPhi} could be due to the parameters that vary between events, such as the solar event magnitude (as measured by the flare class, CME speed, and peak intensity). Solar wind and IMF conditions are expected to play a role as well.

\begin{table*}[h]
\caption{$\tau$ values in hours for the case study events.}
\begin{center}
\small
\begin{tabular}{ |*{5}{c|} }
    \hline 
    \textbf{2021-10-09} & SolO & STEREO-A & SOHO & PSP \\ 
    \hline 
    Electrons $\sim$ 1 MeV  & 3.9 &  4.1 &  15.8 &  7.8 \\
    \hline 
    Protons $\sim$ 25 MeV  & 6.2 & 9.9  & 18.3  &  9.9 \\
    \hline 
    Protons $\sim$ 60 MeV & 3.7 & 4.6  & NaN  &  NaN \\
    \hline 
\end{tabular}
\quad
\begin{tabular}{ |*{5}{c|} }
    \hline 
    \textbf{2021-10-28} & SolO & STEREO-A & SOHO & PSP \\ 
    \hline 
    Electrons $\sim$ 1 MeV  & 21.5 & 16.5  & 24.9  & 19.5  \\
    \hline 
    Protons $\sim$ 25 MeV  & 17.4 &  15.1 & 30.6  &  12.5 \\
    \hline 
    Protons $\sim$ 60 MeV & 13.9 & 12.3  & 17.9  &  7.8 \\
    \hline 
\end{tabular}
\end{center}
\label{tab:CaseStudyTable}
\end{table*}

\subsection{Event with a wide angular separation: 2022 February 15}\label{WideEvent}

\begin{figure}
    \centering
    \includegraphics[width=0.9\linewidth]{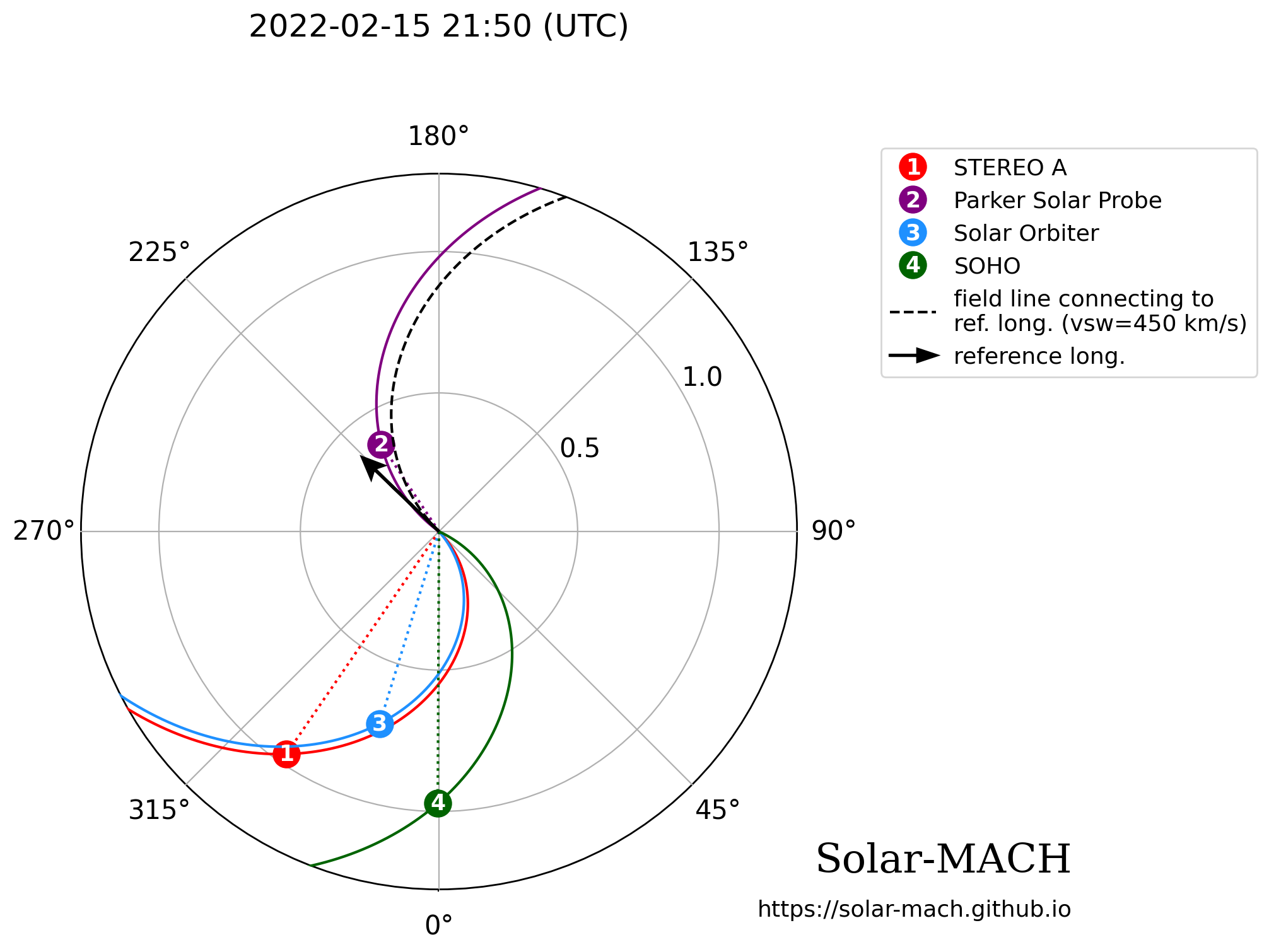}
    
    \caption{Solar-MACH plot for 2022 February 15. The longitudes are in Stonyhurst coordinates. The spacecraft locations and nominal Parker spirals are colour-coded as seen in the key. The Parker spirals were calculated using the solar wind speeds measured by each spacecraft. The black arrow shows the location of the source AR, and the dotted black line shows the nominal Parker spiral for this location, assuming a solar wind speed of 450 km/s.}
    \label{fig:SMExcluded}
\end{figure}

\begin{figure}
    \centering
    \includegraphics[width=1\linewidth]{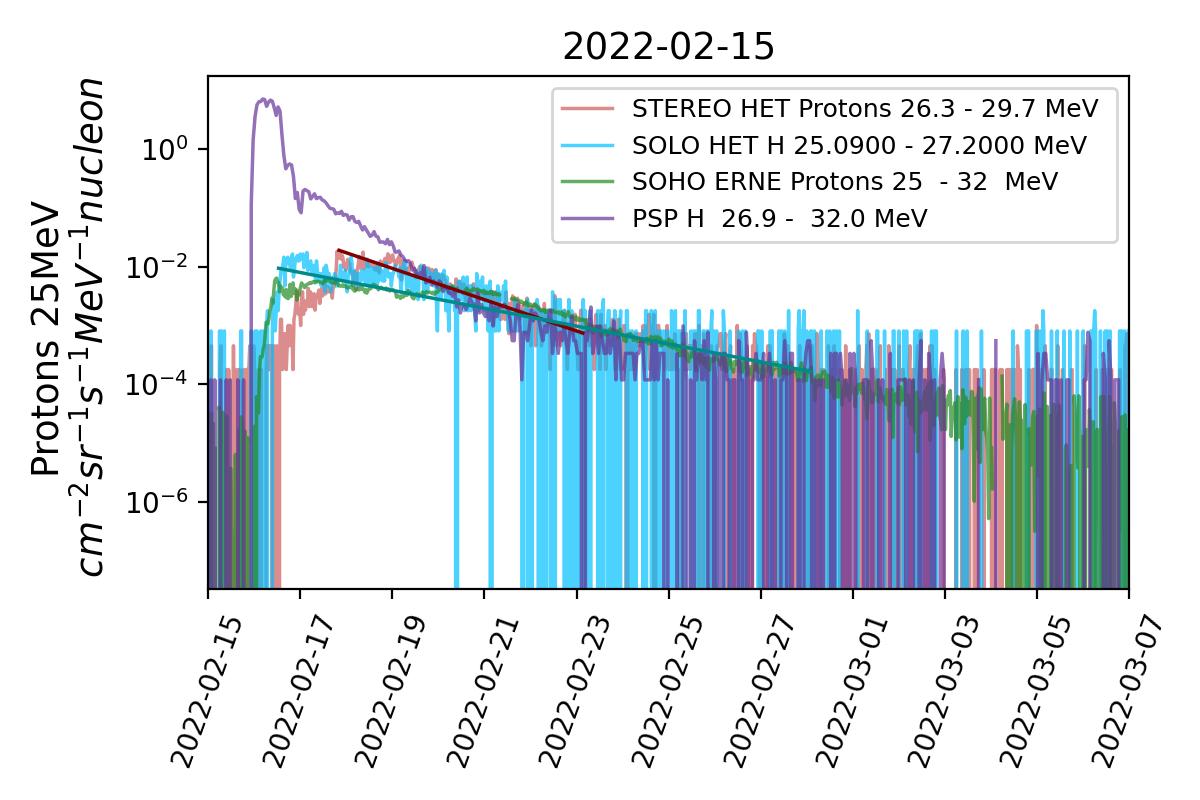}
    \caption{Intensity profiles for the proton $\sim$ 25 MeV channel for the 2022 February 15 event.}
    \label{fig:PlotExcluded}
\end{figure}

An event on 2022 February 15 was studied but not included in Figures \ref{fig:P25TauPhi}-\ref{fig:P60TauPhi}. The flare location and flare time for this event were taken from \citet{khoo_multispacecraft_2024}. Four spacecraft observed the event, and their geometry is shown in Figure \ref{fig:SMExcluded}. PSP was well-connected with $\Delta \phi = -8.5$, but was closer to the Sun than we allowed for our analysis (r = 0.38 AU). The other three spacecraft were connected to the far side of the Sun with respect to the source AR with $\Delta \phi =$ -179.0\textdegree{}, -175.5\textdegree{}, 160.4\textdegree{} for STEREO-A, Solar Orbiter, and SOHO, respectively. 

The profiles for the $\sim$ 25 MeV channel for the four observing spacecraft are shown in Figure \ref{fig:PlotExcluded}. Of the three spacecraft at the far eastern and far western longitude separations, only the STEREO-A and SOHO measurements met our fitting requirements in the $\sim$ 25 MeV proton channel. The STEREO-A $\tau$ value was 39.2 hours and the SOHO $\tau$ value was 68.5 hours. Given that the $\Delta \phi$ values for the spacecraft are -179.0 \textdegree and 160.4 \textdegree, these $\tau$ values would place points beyond the top left and top right corners of Figure \ref{fig:P25TauPhi} respectively. The SOHO $\tau$ value in particular may be affected by fluctuations around the peak value, as shown in Figure \ref{fig:PlotExcluded}.

Figure \ref{fig:PlotExcluded} shows that for these events with a wide angular separation, the rise and peak phase extend over several days, such that the uncertainty in the values of $\tau$ is large. We note that the $\tau$ values in this event are much higher than those in Figure \ref{fig:P25TauPhi}, and the value at SOHO is more than twice the maximum $\tau$ value of any other event. In addition, we only have two points for this event, both at wide longitudinal separations, because PSP was too close to the Sun to be included in our analysis. For these reasons, the event has not been included in Figure \ref{fig:P25TauPhi}.

\subsection{Decay-time constant against proxies of the solar event magnitude}

\begin{figure}
   \centering
   \includegraphics[width=0.95\linewidth]{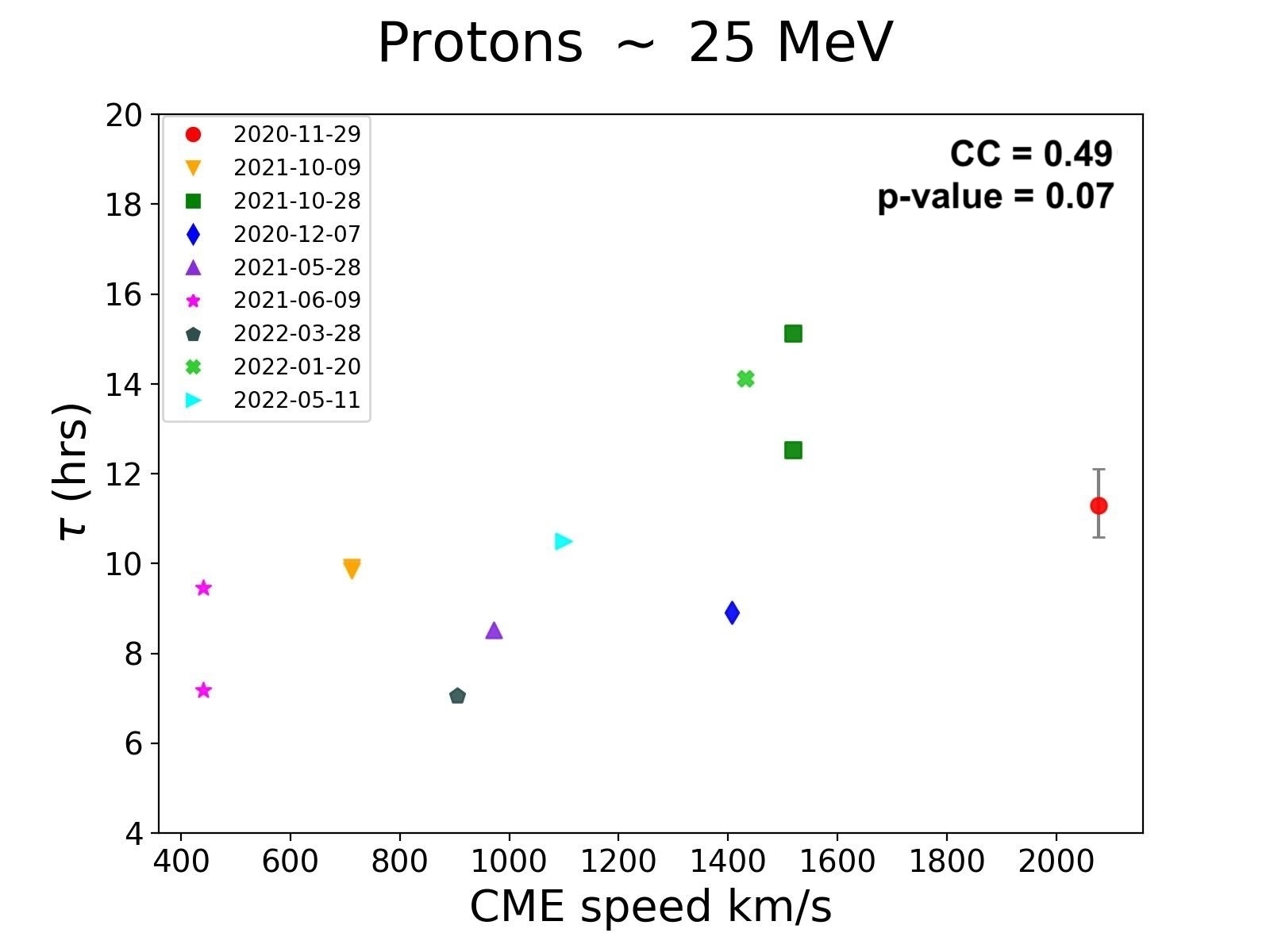}
      \caption{Decay-time constant against CME speed for protons $\sim$ 25 MeV, for all spacecraft observations within $| \Delta \phi | <$  35$^{\circ}$. Where two spacecraft satisfied this condition, both data points are plotted for the same event. The correlation coefficient is CC = 0.49 with a p value = 0.07.
              }
         \label{fig:CMESpeed}
\end{figure}

\begin{figure}
   \centering
   \includegraphics[width=0.95\linewidth]{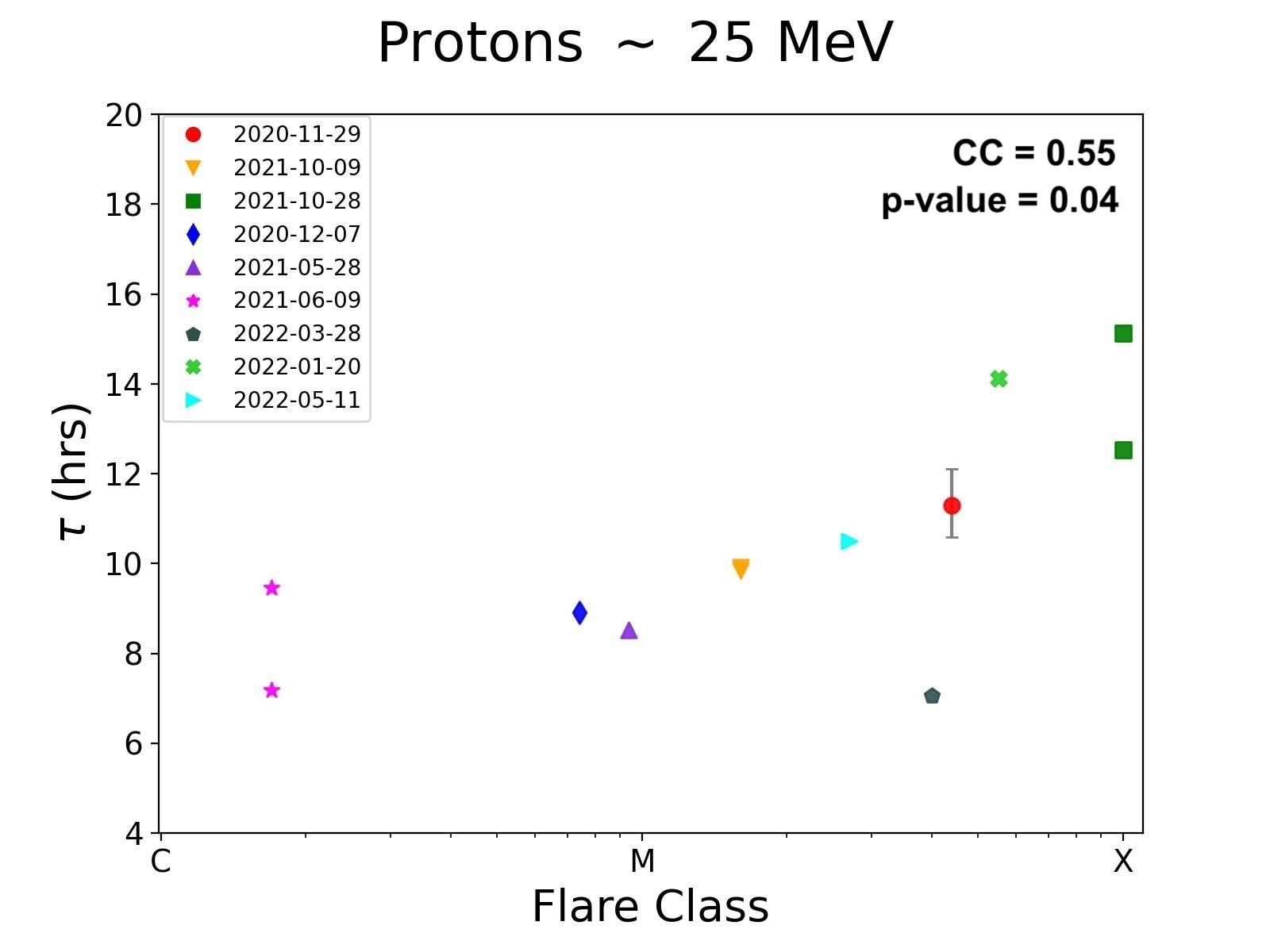}
      \caption{Decay-time constant against GOES soft x-ray flare class for protons $\sim$ 25 MeV, for all spacecraft observations within $| \Delta \phi | <$  35$^{\circ}$. Where two spacecraft satisfied this condition, both data points are plotted for the same event. The correlation coefficient for $\tau$ vs the logarithm of flare class is CC = 0.55, with a p value = 0.04.
              }
         \label{fig:FlareClass}
\end{figure}

\begin{figure}
    \centering
    \includegraphics[width=0.95\linewidth]{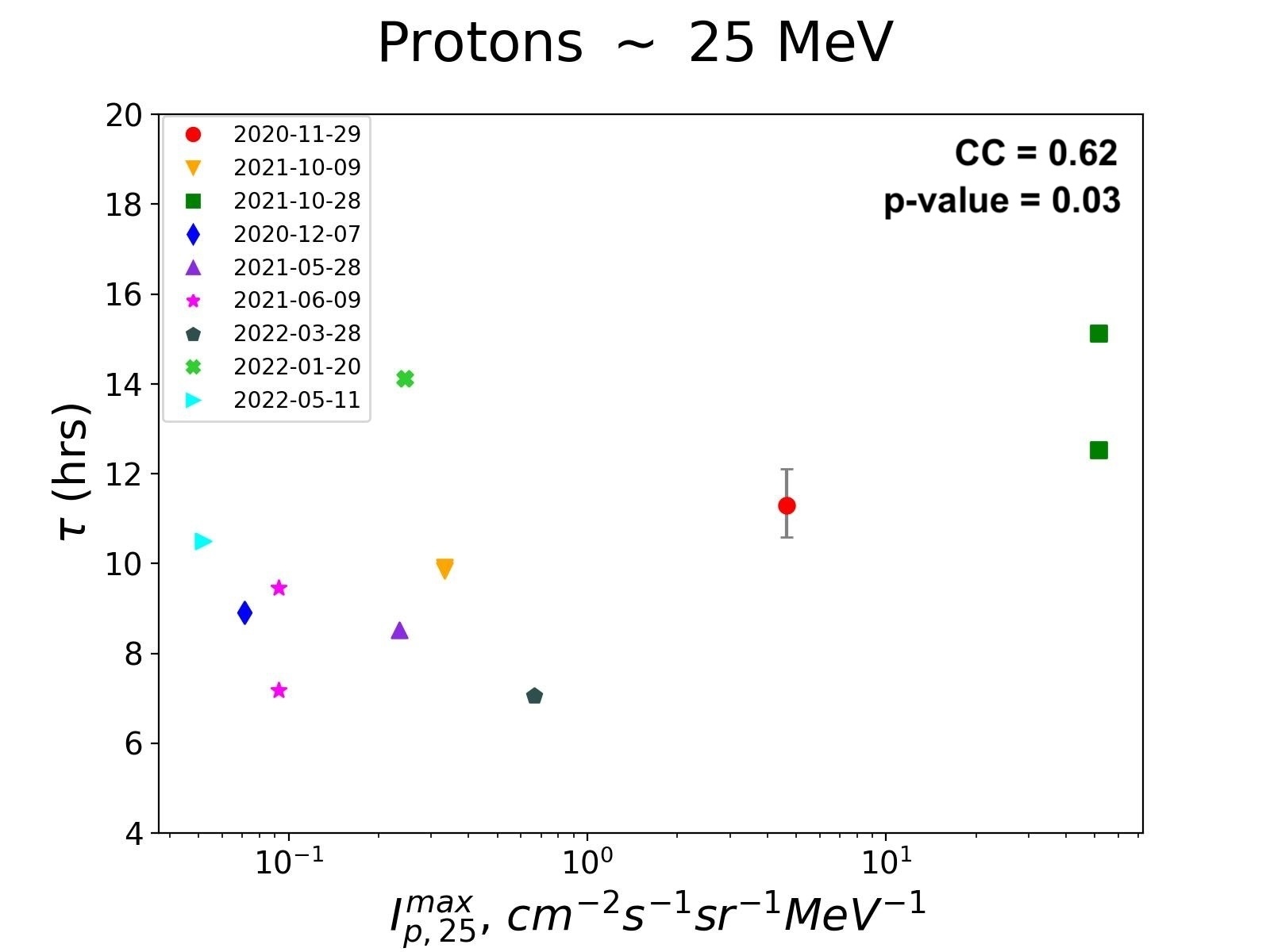}
    \caption{Decay-time constant against $I_{p,25}^{max}$, the maximum proton $\sim$ 25 MeV SEP peak flux over all spacecraft within $| \Delta \phi | <$  35$^{\circ}$. The correlation coefficient for $\tau$ vs the logarithm of $I_{p,25}^{max}$ is CC = 0.62, with a p value = 0.03.
    }
    \label{fig:PeakFlux}
\end{figure}

In an effort to understand the effect that event magnitudes may have on the decay phases, we produced plots of $\tau$ against CME speed, flare class, and SEP peak flux, $I_{p,25}^{max}$. These are shown in Figures \ref{fig:CMESpeed}, \ref{fig:FlareClass}, and \ref{fig:PeakFlux}. In Figure \ref{fig:CMESpeed}, the CME speed is taken as the LASCO POS speed for the associated CME for each event. Where multiple CMEs are associated with the event, the CME with the highest LASCO POS speed was used. Points were used when the spacecraft that took the particle measurements was connected within 35\textdegree{} to the AR associated with the event. The constraint $|\Delta \phi|$ < 35\textdegree{} was used to limit longitudinal effects in the plot and was used in Figures \ref{fig:FlareClass} and \ref{fig:PeakFlux} as well. Only a weak correlation between CME speed and $\tau$ is seen (Pearson correlation coefficient, CC = 0.49 with a p value = 0.07). A faster CME may accelerate more particles over longer times resulting in long-duration decays. It may accelerate particles to higher energies and lengthen the decays as high-energy particles decelerate to fill lower-energy channels over time. The spatial extent over which energetic particles are found may be wider, and combined with corotation, this may lengthen the decay.

In Figure \ref{fig:FlareClass}, $\tau$ is plotted against the GOES soft X-ray flare class of each event. Calculating the correlation coefficient between $\tau$ and logarithm of flare class, we obtain CC = 0.55 with a p value = 0.04. 

In Figure \ref{fig:PeakFlux}, the SEP peak flux, $I_{p,25}^{max}$, is the highest peak event flux over all observing spacecraft within 35 \textdegree in the proton $\sim$ 25 MeV channel. For all our events, we found that the maximum $I_{p,25}$ is associated with the closest connected spacecraft, having the minimum $\Delta \phi$ value. The proton $\sim$ 25 MeV channel was chosen because it overlaps most for the spacecraft instruments and is most reliable for obtaining measurements, regardless of the event magnitude. 
There is a weak trend for an increase in $\tau$ as $I_{p,25}^{max}$ increases (CC = 0.62 with a p value = 0.03). Nine out of 12 of the points are on the left side of the graph, with $I_{p,25}^{max}$ lower than $1$ \textrm{cm}$^{-2}$\textrm{s}$^{-1}$\textrm{sr}$^{-1}$\textrm{MeV}$^{-1}$. This leaves a cluster of points on the left side, and a sparse set of points on the right side. The high-peak intensity points on the right side of the graph strongly affect the correlation coefficient. If the two green points were removed for 2021 October 28, there would be no correlation.

\subsection{Fitting late-phase decay versus the entire decay}\label{LarioDecay}

The decay-time constant values presented in Section \ref{results} were derived by fitting intensities between 90\% of $I_p$ and the time at which the intensities again reached background levels. However, other choices are possible for the fit. \cite{lario_statistical_2010} identified two phases in the decay of several events observed during solar cycle 23. In these events, which tended to be well-connected events, they observed an initial rapid power-law decay phase, and a later exponential decay phase. They chose to fit only the later exponential decay phase when they derived $\tau$, which generally results in higher values of $\tau$ for these events compared to when the earlier decay phases are included.

\begin{figure}
    \centering
    \includegraphics[width=0.95\linewidth]{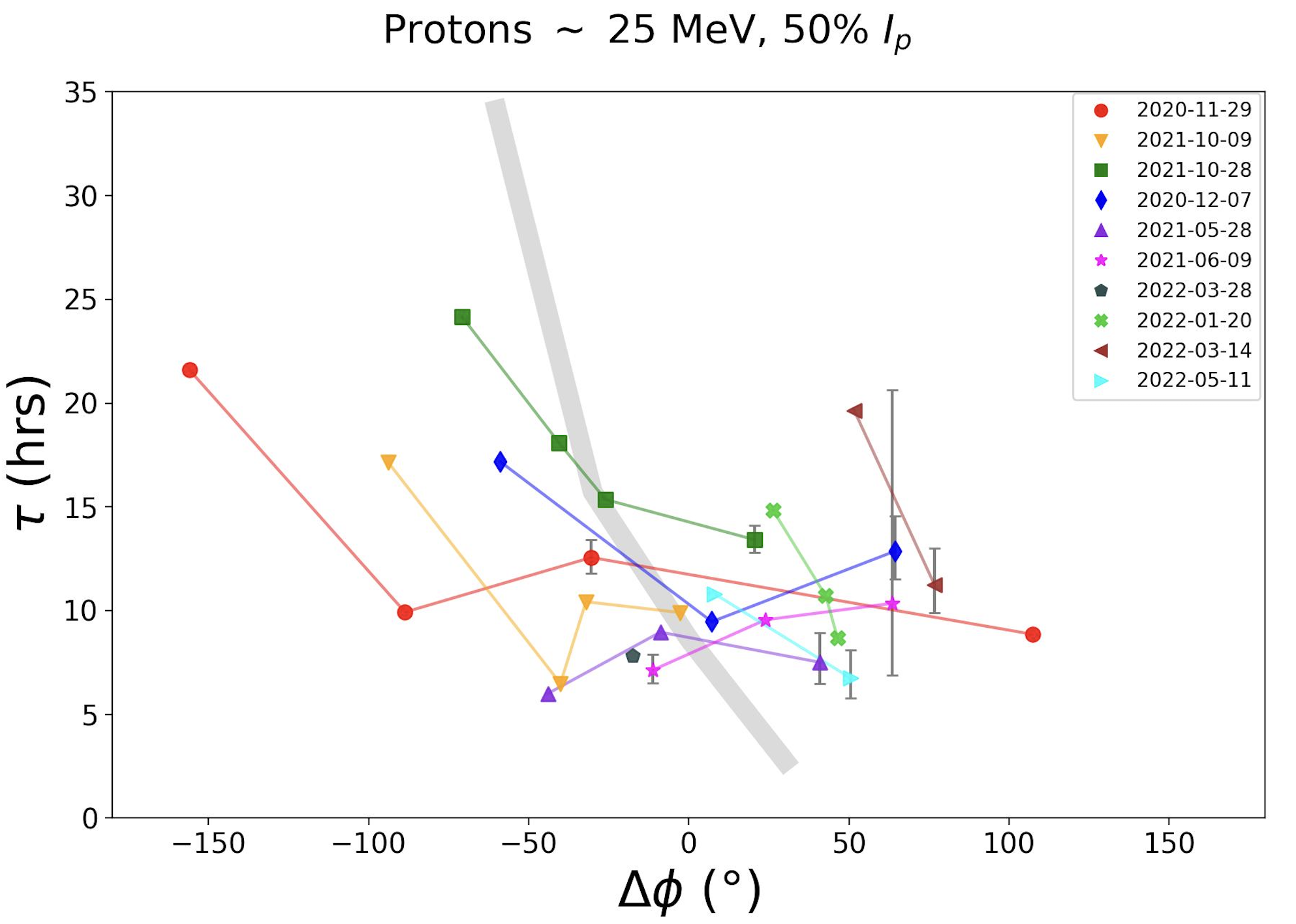}
    \caption{Decay-time constant $\tau$ vs. longitudinal separation $\Delta \phi$ (as given by Equation \ref{eqn:DPEqn}) for $\sim$ 25 MeV protons. The start of the decay phase is defined as when intensities return to 50\% of $I_p$. The coloured lines connect s/c data points for a single event. The grey shading shows results from a 25 MeV simulation run following the method of \citet{hutchinson_modelling_2023}. The error bars were omitted when they are smaller than the data points.}
    \label{fig:0.5ThreeChannels}
\end{figure}

\begin{figure}
    \centering
    \includegraphics[width=\linewidth]{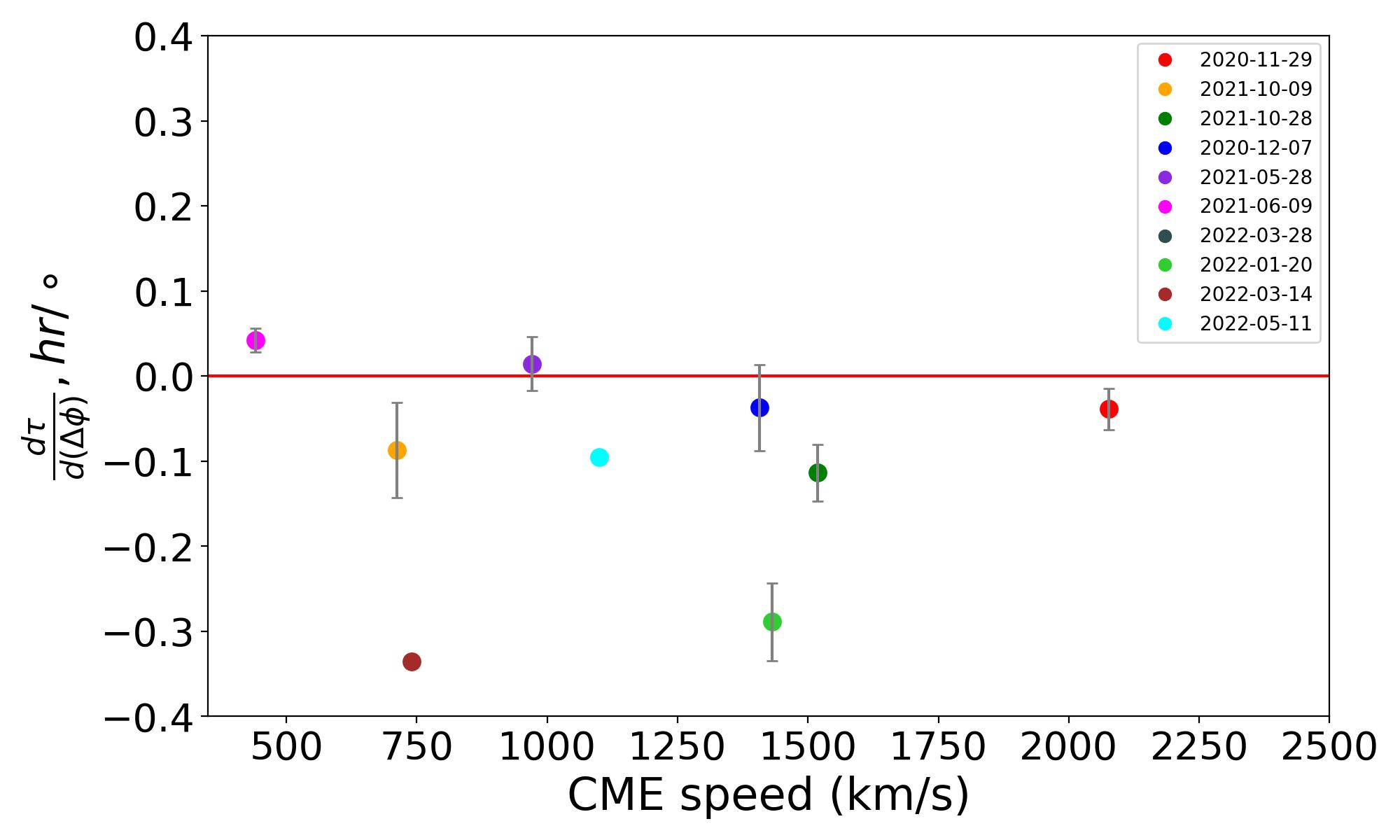}
    \caption{Slope values from the linear fit of $\tau$ vs. $\Delta \phi$ data points for each event against CME speed. The start of the decay phase is defined as when intensities return to 50\% of $I_p$.}
    \label{fig:0.5CME}
\end{figure}

We analysed the effect when only the later phase of the event is fitted by fitting intensities between 50\% of $I_p$ and the time when background is reached. Figure \ref{fig:0.5ThreeChannels} shows the effect of this choice to be compared with Figure \ref{fig:P25TauPhi}. There are some differences compared to Figure \ref{fig:P25TauPhi} but the main trends remain the same. The errors on the $\tau$ values are also larger in Figure \ref{fig:0.5ThreeChannels} because fewer data points are included. Figure \ref{fig:0.5CME} shows the result of recreating Figure \ref{fig:SlopeCME} using 50\% of $I_p$ instead of 90\% of $I_p$, and the majority of the events clearly follow the trend of decreasing $\tau$ values with increasingly western $\Delta \phi$ values. The two events with positive $\frac{d\tau}{d(\Delta \phi)}$ values in Figure \ref{fig:SlopeCME} have slightly higher positive $\frac{d\tau}{d(\Delta \phi)}$ values in Figure \ref{fig:0.5CME}, but are still in the minority. This shows that the east-west trend we see is not simply caused by the presence of an initial faster decay phase seen by observers with a closer connection to the acceleration region of the SEP event.

\section{Discussion and conclusions} 

We have analysed 11 SEP events with four observing spacecraft, with a focus on the decay phase. We determined the decay-time constant, $\tau$, in two proton channels and one electron channel and studied its dependence on $\Delta \phi$, which is the longitudinal separation between the source AR and observer footpoint location at the Sun.

The main results of this work are listed below.

   \begin{enumerate}
      \item Within individual events there is a trend of decreasing $\tau$ values for increasingly western ARs ($\tau$ decreasing with increasing $\Delta \phi$ (Figures \ref{fig:P25TauPhi} and \ref{fig:SlopeCME})). This is seen for both electrons and protons, but more data points for higher-energy protons are needed to investigate the consistency of the trend at different proton energies.
      \item The east-west trend for individual events is present regardless of whether we fit the whole decay phase or only a later stage of the decay phases (Figure \ref{fig:0.5ThreeChannels}). 
      \item 
      The overall magnitude of the events affects the value of $\tau$  (Section \ref{CaseStudySection}), and it is likely that transport conditions also play a role. There are weak trends for $\tau$ to increase with CME speed (CC = 0.49) and flare class (CC = 0.55), and a slightly stronger trend for $\tau$ to increase with SEP peak intensity (CC = 0.62). (Figures \ref{fig:CMESpeed} to \ref{fig:PeakFlux}). Nine events were included in this analysis.

   \end{enumerate}

The multi-spacecraft element of the analysis has been key to identifying a dependence of $\tau$ on $\Delta \phi$. \citet{lario_statistical_2010} did not find a systematic dependence of $\tau$ on longitude. This may be because each point in their plot corresponds to a separate event and many parameters vary from event to event. For example, the solar event magnitude (as discussed in Section \ref{CaseStudySection}) as well as the solar wind and IMF conditions play an important role in determining the $\tau$ value. For their plot of $\tau$ against longitude (their Figure 10), they used the longitude of the source AR as their X-axis. Thus they did not take possible effects of the solar wind speed on the magnetic connection of the observer into account. Our conclusion is that when only single-spacecraft events are included in a study, the east-west trend is hidden by the large variability in $\tau$ values in different events. In our analysis, we tried to derive $\Delta \phi$ values for the events as accurately as possible, but these trends may be influenced by factors such as turbulence, field line meandering and coronal and interplanetary structures \citep{wimmer-schweingruber_unusually_2023}. 

\cite{dalla_multisc_2003} used data from Helios 1 and 2 and IMP8 to study the dependence of the event duration on $\Delta \phi$ for 52 gradual events. They found a trend for the longest durations to be associated with events with large negative $\Delta \phi$ and for events with large positive $\Delta \phi$ to have short durations. Since there is a correlation between duration and decay-time constant, their results are consistent with those presented in Figure \ref{fig:P25TauPhi}.

In addition to the event magnitude discussed above, a number of other influences on decay-time constants are likely to play a role. Previous studies have found effects of the radial distance of the observing spacecraft from the Sun on the decay phases. \citet{kecskemety_decay_2009} compared decay-time constants for events at Helios 1 and 2 and IMP to those at Ulysses, covering radial ranges of $\sim$ 1-5 AU. 
They found that $\tau$ increased with increasing radial distance between 2-5 AU. A similar trend was found by \citet{lario_statistical_2010}. 
In our analysis, in order to separate the observer longitude effects from observer radial effects, we chose to only include measurements of SEPs from spacecraft when they were located farther away than 0.6 AU from the Sun. 

The east-west trend in the $\tau$ values can be interpreted as a signature of corotation, guided by the results of simulations by \citet{hutchinson_impact_2023} who showed that corotation introduces a systematic decrease in $\tau$ with increasing $\Delta \phi$ compared to simulations that do not include corotation. An alternative possibility is that this trend is due to the temporal and spatial dependence of the particle acceleration. In particular it has been suggested that the variation in the profile parameters with $\Delta \phi$ may be due to changes in the acceleration efficiency along a CME-driven shock front that accelerates the particles \citep{reames_spatial_1996}. The way in which the energetic particle profiles at different observers are affected by the CME shock properties was discussed for example by \citet{hu_modeling_2017}.
Finally, the different interplanetary transport conditions at different longitudes may also explain the different $\tau$ values, but not the systematic decrease with $\Delta \phi$.

\citet{hutchinson_modelling_2023} modelled a time-extended injection from a broad shock in the presence of corotation and showed that the measured intensity profiles at different observers depend only weakly on the characteristic of the injection at the shock (e.g. on whether the injection at the shock is Gaussian or uniform). It is interesting to note that in the past, the fitting of the decay phase was used to determine the value of the scattering mean free path $\lambda$ within 1D transport models. \citet{hutchinson_modelling_2023} also showed that when corotation is included, the decay-time constant shows little dependence on the value of the scattering mean free path.

The simulations of \citet{hutchinson_impact_2023} did not include the effects of perpendicular transport on SEPs. By enabling particles to propagate across magnetic flux tubes as time goes on, perpendicular transport would be expected to reduce the signatures of corotation in the decay phase. This may explain why in the majority of events we studied, the gradient $\frac{d\tau}{d(\Delta \phi)}$ is less steep than in the simulations.

The test-particle simulations in \citet{hutchinson_impact_2023} indicated that corotation can be important in shaping the SEP decay phases and should be included in SEP models and in interpretations of events. The systematic nature of the $\tau$ versus $\Delta \phi$ slopes observed in the events that we analysed appears to support the simulation results. In a separate study, \cite{dalla_detection_2024} analysed the distribution of the SEP event occurrence in $\Delta \phi$ and showed an asymmetry in the detection, which might also be a signature of corotation. We conclude that corotation should be included in SEP models and in interpretations of events, although other processes such as acceleration and transport may be involved in producing the observed trends. 

Future work should include a larger sample of events. This will be made possible as solar cycle 25 continues and multi-spacecraft measurements continue to be taken. Future studies with a larger sample of events could also group events with similar parameters such as flare class and SEP peak flux. This may aid in reducing the spread of $\tau$ values we observed and might clarify an east-west trend when events are viewed together.

\section{Data availability}
All spacecraft data used in this paper are publicly available and can be retrieved using the EU SERPENTINE software \citep{palmroos_solar_2022}. Details of the events studied are provided in Table \ref{tab:Events}. Results derived from our data analysis are available in Table \ref{tab:long}.

\begin{acknowledgements}
R.H. acknowledges funding from the Moses Holden Studentship for her PhD. 

T.L. and S.D. acknowledge support from the UK Science and Technology Facilities Council (STFC) through grants ST/V000934/1 and ST/Y002725/1. 

A.H. would like to acknowledge support from the University of Maryland Baltimore County (UMBC), the Partnership for Heliophysics and Space Environment Research (PHaSER), and NASA/GFSC.

We acknowledge use of solar energetic particle data from the SOHO, STEREO-A, Solar Orbiter and PSP spacecraft and thank the instrument teams for their work on making the data available and science-ready. 

Solar Orbiter is a mission of international cooperation between ESA and NASA, operated by ESA.
Thanks to the Integrated Science Investigation of the Sun (IS$\odot$IS) Science Team (PI: David McComas, Princeton University), and the Energetic Particle Detector (EPD) Team (PI: Javier
Rodríguez-Pacheco, University of Alcalá, Spain).

We acknowledge use of SERPENTINE tools, which were developed with funding from the European Union's Horizon 2020 research and innovation program, and of the Solar-MACH tool. 

The use of the data made available via the NSSDC CDAWeb is acknowledged.

\end{acknowledgements}

\begin{appendix}
\onecolumn
\section{Further event details}
\begin{longtable}{|c||c|c|c|c|c|c|c|c|}
\caption{Event parameters and results.}
\label{tab:long} \\

\hline
Event Date & Spacecraft & $\Delta \phi$, \textdegree & \shortstack{\\Radial distance, \\ au} & $\tau_{E1}$& $\tau_{P25}$& $\tau_{P60}$ & $\frac{\mathrm{d} \tau}{\mathrm{d}(\Delta \phi)}$, hrs/\textdegree &  Error on $\frac{\mathrm{d} \tau}{\mathrm{d}(\Delta \phi)}$ \\ 
\endfirsthead

\multicolumn{9}{c}%
{{\bfseries \tablename\ \thetable{} -- continued from previous page}} \\
Event Date & Spacecraft & $\Delta \phi$, \textdegree & \shortstack{\\Radial distance, \\ au} & $\tau_{E1}$& $\tau_{P25}$& $\tau_{P60}$ & $\frac{\mathrm{d} \tau}{\mathrm{d}(\Delta \phi)}$, hrs/\textdegree &  Error on $\frac{\mathrm{d} \tau}{\mathrm{d}(\Delta \phi)}$ \\ 
\endhead

\hline \multicolumn{9}{|r|}{{Continued on next page}} \\ \hline
\endfoot

\hline \hline
\endlastfoot
    \hline 
29/11/2020	& SOHO		& -156.0		& 0.98		&  	30.8		& 22.7	& 	- & -0.0429    		& 0.0265	
\\  
    \hline 
-		& STEREO-A		& -88.6		& 0.96		& 12.3		& 10.2		& - & -		& -		
\\ 
    \hline 
-		& PSP		& -30.4	& 	0.81	& 	15.3	& 	11.3	& 	- & 	-		& -		
\\ 
    \hline 
-		& SolO		& 107.4		& 0.88	& 15.6	& 	8.9		& 6.6 & 	-	& 	-			
\\ 
    \hline 
07/12/2020		& SOHO		& -58.9	& 	0.98	&	-		& 19.0	& 	-	& -0.0526    	& 	0.0638	 
\\ 
    \hline 
-	& 	STEREO-A		& 7.2	& 	0.96	& -	& 	 8.9		& 4.5 & 	-	& 	-		
\\ 
    \hline 
-	& 	PSP	& 	64.5		& 0.78	& -	& 12.9		& - & 	-	& 	-		
\\ 
    \hline 
-		& SolO		& -161.0		& 0.84	& 	-	& 	-		& -& 	-	& 	-		
\\ 
    \hline 
28/05/2021		& SOHO		& -8.9	& 	1.00		& 5.7		& 8.5	& 	- & 0.0089		& 0.0293 			
\\ 
    \hline 
-		& STEREO-A	& 	41.0		& 0.96	& -		& 6.9	& 	-	& -		& -			
\\ 
    \hline 
-	& 	PSP		& -44.0		& 0.69	& 8.2	& 	5.9		& -	& -		& -			
\\ 
    \hline 
-	& 	SolO		& 87.8		& 0.95	& -		& -		& -	& -			& -		
\\ 
    \hline 
09/06/2021	& 	SOHO	& 	23.9	& 	1.01	& 6.6		& 9.5		& -	& 0.0150 		& 0.0267 			
\\ 
    \hline 
-	& 	STEREO-A		& 63.6	& 	0.96	& 	-		& 8.4		& - & 	-	& 	-	
\\ 
    \hline 
-	& 	PSP		& -11.2		& 0.76	& -		& 7.2		& -	& -		& -			
\\ 
    \hline 
-		& SolO		& 135.7	& 	0.95	&-		& -		& - & 	-	& 	-	
\\ 
    \hline 
09/10/2021	& 	SOHO		& -93.9		& 0.99	& 	15.8		& 18.3		& -	& -0.1019   	& 	0.0624 		
\\ 
    \hline 
-		& STEREO-A		& -32.1		& 0.96	& 4.1	& 	9.9		& 4.6	& -		& -			
\\ 
    \hline 
-		& PSP		& -2.8	& 	0.77	& 	7.7		& 9.9		& - & 	-	& 	-	
\\ 
    \hline 
-		& SolO		& -40.1		& 0.68	& 3.9		& 6.2	& 	3.7 & 	-		& -			
\\ 
    \hline 
28/10/2021	& 	SOHO		& -70.8	& 	0.98		& 	24.9		& 30.6	& 	17.9 & -0.1836   		& 0.0737 		
\\ 
    \hline 
-		& STEREO-A		& -26.1		& 0.96 & 	 16.5	& 	15.1	& 	12.3	& 	-	& 	-	
\\ 
    \hline 
-		& PSP		& 20.5	& 	0.62	& 19.5	& 	12.5		& - & 	-		& -			
\\ 
    \hline 
-		& SolO		& -40.5		& 0.80	& 21.5		& 17.4		& 13.9	& -	& 	-			
\\ \hline
20/01/2022		& SOHO		& 26.3	& 0.98	& 14.3	& 14.1	& 5.7 & -0.2783   & 	0.0585 			
\\ 
    \hline 
-		& STEREO-A	& 	42.8	& 	0.97	& 8.8 & 10.3 & 7.1 & 	-	& 	-			
\\ 
    \hline 
-		& PSP		& -177.6		& 0.73	& -		& -		& -	& -		& -			
\\ 
    \hline 
-	& 	SolO	& 	46.4	& 	0.92	& 	7.9	& 8.1	& 	6.1 & 	-	& 	-	
\\ 
    \hline 
15/02/2022		& SOHO		& 160.4	& 	0.98	& 	-			& 68.5		& - & 	-	& 	-	
\\ 
    \hline 
-	& 	STEREO-A	& 	-179.0		& 0.97	& 38.6	& 39.2	& 	32.7 & 	-	& 		-		
\\ 
    \hline 
-	& 	PSP		& -8.5	& 	0.38 & 	-	& -		& -	& 	-	& 	-	
\\ 
    \hline 
-		& SolO		& -175.5		& 0.72 & 53.4		& -		& -	& 	-	& 	-			
\\ 
    \hline 
14/03/2022		& SOHO		& 51.6	& 	0.98	& 8.6 & 19.8	& - & -0.3570   		& 0.0 			
\\ 
    \hline 
-		& STEREO-A		& 76.7	& 	0.97	& -		& 10.8	& 	- & 	-		& -			
\\ 
    \hline 
-	& 	PSP		& -49.2		& 0.53	& -		& -		& - & 	-	& 	-		
\\ 
    \hline 
-	& 	SolO		& 64.6		& 0.41 & -		& -		& -	& 	-	& 	-			
\\ 
    \hline 
28/03/2022	& 	SOHO		& -47.6		& 0.99 & 7.4 & 10.9		& 6.3		& -0.1269 	& 0.0 		
\\ 
    \hline 
-	& 	STEREO-A		& -17.5		& 0.97 & 5.6 & 7.1	& 	5.4	& 	-	& 	-			
\\ 
    \hline 
-	& 	PSP		& -166.7		& 0.69 & 	-			& -		& -	& 	-	& 	-	
\\ 
    \hline 
-	& 	SolO	& 	-111.6		& 0.33 & 	-			& -		& - & 	-	& 	-	
\\ 
    \hline 
11/05/2022	& 	SOHO		& 8.0		& 1.00	& 8.6		& 10.5	& 	-	& -0.1146   	& 	0.0 		
\\ 
    \hline 
-	& 	STEREO-A		& 50.6	& 	0.96	& 	-			& 5.6		& -	& -		& -	
\\ 
    \hline 
-	& 	PSP		& -67.6		& 0.59	& -		& -		& -	& -		& -			
\\ 
    \hline 
-	& 	SolO		& -128.2		& 0.79	& -		& -		& - & 	-		& -		
\\ 
    \hline 
    
\end{longtable}
\tablefoot{Event dates are taken from the SERPENTINE Events Catalog \citep{dresing_serpentine_2024}. Spacecraft location is given as calculated using Solar-MACH \citep{gieseler_solar-mach_2023}. The $\tau$ values found for each particle channel as well as the $\frac{\mathrm{d} \tau}{\mathrm{d}(\Delta \phi)}$ values (for the $\sim$ 25 MeV proton channel) with their errors are also given.}

\end{appendix}

\end{document}